\documentclass{pr-imfp03-new}

\usepackage{amssymb}

\begin{document}

\begin{flushright}
DESY-04-036\\
ZU-TH 02/04\\
March 2004\\
\end{flushright}

\vspace*{1.5cm}
\begin{center}
    {\baselineskip 25pt
    \Large{\bf

An Analysis of the Time-Dependent CP Asymmetry \\ 
in $B \to \pi \pi$  Decays in the Standard Model

    }
    }

\vspace{1.2cm}
\centerline{\bf Ahmed Ali}

\vspace{.5cm}
\small{\it Theory Group,
 Deutsches Elektronen-Synchrotron DESY,
 D-22603 Hamburg, FRG.
\footnote{E-mail: ahmed.ali@desy.de}}
\vspace{.5cm}

\centerline{\bf Enrico Lunghi}
\vspace{.5cm}
\small{\it Institut f\"ur Theoretische Physik, 
 Universit\"at Z\"urich, 
 CH-8057 Z\"urich, Switzerland.
\footnote{E-mail: lunghi@physik.unizh.ch}}

\vspace{.5cm}
\centerline{\bf Alexander Ya.~Parkhomenko}
\vspace{.5cm}
\small{\it Institut f\"ur Theoretische Physik, 
 Universit\"at Bern, 
 CH-3012 Bern, Switzerland.
\footnote{E-mail: parkh@itp.unibe.ch} 
\footnote{On leave of absence from 
Department of Theoretical Physics,
Yaroslavl State University, Sovietskaya 14, 
150000 Yaroslavl, Russia.}}  

    \vspace{.5cm}
    \today

    \vspace{1.5cm}
    {\bf Abstract}

\end{center}

\bigskip

Measurements of the time-dependent CP asymmetry in the decay
$B^0_d (t)\to \pi^+ \pi^-$ and its charge conjugate 
by the BELLE and BABAR collaborations currently yield 
$C_{\pi \pi}^{+-} = -0.46 \pm 0.13$ and 
$S_{\pi \pi}^{+-} = -0.74 \pm  0.16$, characterizing 
the direct and mixing-induced CP asymmetries, respectively. 
We study the implication of these measurements on the CKM 
phenomenology taking into account the available information 
in the quark mixing sector. Our analysis leads to the results 
that the ratio~$|P_c/T_c|$ involving the QCD-penguin and 
tree amplitudes and the related strong phase difference 
$\delta_c = \delta_c^P - \delta_c^T$ in the 
$B^0_d/\bar{B}_d^0 \to \pi^+ \pi^-$ decays are quite substantial. 
Using the isospin symmetry to constrain~$|P_c/T_c|$ and 
$\cos (2\theta)$, where~$2 \theta$ parameterizes the penguin-induced 
contribution, we present a fit of the current data
including the measurements of~$S_{\pi \pi}^{+-}$ and~$C_{\pi \pi}^{+-}$.
Our best-fits yield: $\alpha = 92^\circ$, $\beta = 24^\circ$, 
$\gamma = 64^\circ$, $\vert P_c/T_c\vert = 0.77$, and 
$\delta_c = -43^\circ$. At 68\%~C.L., the ranges are:
$81^\circ \leq \alpha \leq 103^\circ$, 
$21.9^\circ \leq \beta \leq 25.5^\circ$,
$54^\circ \leq \gamma \leq 75^\circ$, 
$0.43 \leq \vert P_c/T_c\vert \leq 1.35$ and
$-64^\circ \leq \delta_c \leq -29^\circ$. 
Currently {\it en vogue} dynamical approaches to estimate the hadronic
matrix elements in $B \to \pi\pi$ decays do not provide a good fit of
the current data.

\newpage


\section{Introduction}
\begin{figure}[tbp]
%
\centerline{\psfig{width=0.45\textwidth,file=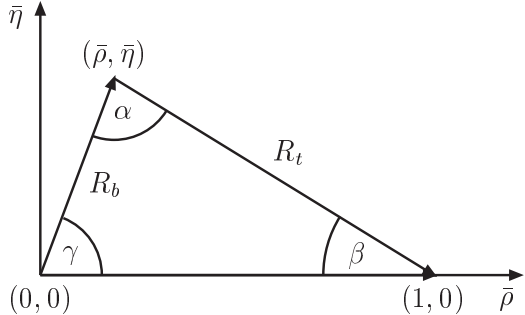}}
\caption{The unitarity triangle with the unit base in  
   the $\bar \rho - \bar \eta$ plane. The two sides~$R_b$ 
   and~$R_t$ and the angles~$\alpha$ ($\phi_2$), $\beta$ 
   ($\phi_1$) and~$\gamma$ ($\phi_3$) are defined in 
   Eqs.~(\ref{eq:Rb-Rt}) and~(\ref{eq:abgamma}), respectively.} 
\label{fig:triangle}
\end{figure}

Precise measurement of CP-violation in $B$-meson decays is  
the principal goal of experiments at the current electron-positron 
$B$-factories, KEK-B and SLAC-B, and at the  hadron colliders, 
Tevatron and LHC. In the standard model (SM), the source of 
CP violation is the Kobayashi-Maskawa phase~\cite{Kobayashi:fv}
which resides in the Cabibbo-Kobayashi-Maskawa (CKM)
matrix~\cite{Kobayashi:fv,Cabibbo:yz}.
In the Wolfenstein parameterization~\cite{Wolfenstein:1983yz} of the 
CKM matrix, characterized by the parameters~$\lambda$, $A$, $\rho$ 
and~$\eta$, CP violation is related to a non-zero value of the 
parameter~$\eta$. Of particular importance in the analysis of  
CP violation in the $B$-meson sector is the following unitarity
relation:    
\begin{equation}
V_{ud} V_{ub}^* + V_{cd} V_{cb}^* + V_{td} V_{tb}^* = 0 ,
\label{eq:vudtriangle}
\end{equation}
which is a triangle relation in the complex 
$\bar{\rho} - \bar{\eta}$ plane, depicted in Fig.~\ref{fig:triangle}.
Here, $\bar \rho = (1 - \lambda^2/2) \, \rho$
and $\bar \eta = (1 - \lambda^2/2) \, \eta$ are the perturbatively
improved Wolfenstein parameters~\cite{Buras:1994ec}. The sides of 
this triangle, called~$R_b$ and~$R_t$, are defined as 
\begin{equation}
R_b \equiv \sqrt{\bar \rho^2 + \bar \eta^2},  
\qquad  
R_t \equiv \sqrt{(1 - \bar \rho)^2 + \bar \eta^2},
\label{eq:Rb-Rt}
\end{equation}
and its three inner angles 
have their usual definitions:
\begin{equation}
\alpha \equiv \arg
\left( -\frac{V_{tb}^* V_{td}}{V_{ub}^* V_{ud}} \right) \,,
\quad
\beta \equiv \arg
\left( -\frac{V_{cb}^* V_{cd}}{V_{tb}^* V_{td}} \right) \,,
\quad
\gamma \equiv \arg
\left( -\frac{V_{ub}^* V_{ud}}{V_{cb}^* V_{cd}} \right) \,.
\label{eq:abgamma}
\end{equation}
The BELLE collaboration uses a different notation for these angles:
$\phi_1 = \beta$, $\phi_2 = \alpha$, and $\phi_3 = \gamma$.
We recall that among the CKM matrix elements above~$V_{ub}$ 
and~$V_{td}$ have sizable imaginary parts, and hence all three 
angles $\alpha$, $\beta$ and $\gamma$ are sizable. Of these, the 
phase~$\beta$ has already been well measured using the time-dependent 
CP asymmetries in the $B \to J/\psi K_S$ and related decays, 
yielding~\cite{Browder:2003}: 
\begin{equation}
\sin (2\beta) = \sin (2\phi_1) = 0.736 \pm 0.049 ,  
\qquad 
\beta = \phi_1 = 23.8^\circ \pm 2.0^\circ .
\label{eq:sin2beta-exp}
\end{equation} 
The current thrust~\cite{Jawahery:2003} of the two $B$-factory
experiments~-- BABAR and BELLE~-- is now on the measurements of 
the other two angles~$\alpha$ (or $\phi_2$) and~$\gamma$ 
(or $\phi_3$). Of these, the weak phase~$\alpha$ will be measured 
through the CP violation in the $B \to \pi \pi$, $B \to \rho \pi$ 
and $B \to \rho \rho$ decays. To eliminate the hadronic uncertainties 
in the determination of~$\alpha$, an isospin analysis of these final 
states (as well as an angular analysis in 
the~$\rho \rho$ case) will be necessary~\cite{Gronau:1990ka}. 
To carry out the isospin analysis in $B \to \pi \pi$ decays, one needs 
to know the three amplitudes~$A^{+-}$, $A^{00}$ and~$A^{+0}$, 
corresponding to the $B^0_d \to \pi^+ \pi^-$, $B^0_d \to \pi^0 \pi^0$ 
and $B^+\to \pi^+ \pi^0$ decays, respectively, and their charge
conjugates~$\bar{A}^{ij}$. At present, the only missing pieces 
in the current data are~$A^{00}$ and~$\bar A^{00}$ -- the amplitudes 
of the $B^0_d \to \pi^0 \pi^0$ and $\bar B^0_d \to \pi^0 \pi^0$ 
decays, respectively -- though the measured charge conjugate 
averaged branching ratio~\cite{Fry:2003}
${\cal B} (B^0/\overline{B^0} \to \pi^0 \pi^0)$ provides an 
information on the sum $|\bar A^{00}|^2 + |A^{00}|^2$. Hence,
a model-independent isospin analysis of the $B \to \pi \pi$ 
decays can not be carried out at present from the branching 
ratios alone.  

%
%

In addition to the measurements of the branching ratios in the 
$B \to \pi \pi$ decays, the time-dependent CP asymmetries in 
the $B^0_d (t)/\bar{B}^0_d (t) \to \pi^+ \pi^-$ and 
$B^0_d (t)/\bar{B}^0_d (t) \to \pi^0 \pi^0$ decays will greatly 
help in pinning down the weak phase~$\alpha$. We shall concentrate 
here on the CP-asymmetry in the decay $B^0_d \to \pi^+ \pi^-$,
which is defined as follows:
\begin{eqnarray}
a_{\pi \pi}^{+-} (t) & \equiv &
\frac{\Gamma [\bar B^0_d (t) \to \pi^+ \pi^-] -   
      \Gamma [B^0_d (t) \to \pi^+ \pi^-]}
     {\Gamma [\bar B^0_d (t) \to \pi^+ \pi^-] +
      \Gamma [B^0_d (t) \to \pi^+ \pi^-]}
\label{eq:ACP-def} \\[1mm]
& = & S_{\pi \pi}^{+-} \sin (\Delta M_B \, t)
    - C_{\pi \pi}^{+-} \cos (\Delta M_B \, t) , 
\nonumber
\end{eqnarray}
where $\Delta M_B$ is the mass difference in the $B^0_d - \bar B^0_d$ 
system which is already well measured~\cite{HFAG:2004}, 
and~$C_{\pi \pi}^{+-}$  and~$S_{\pi \pi}^{+-}$ are the direct 
and mixing-induced CP asymmetry parameters, respectively.    
In the notation used by the BELLE collaboration~\cite{Abe:2003ja}, 
$C_{\pi \pi}^{+-}$ is replaced by $A_{\pi \pi}^{+-}$, where 
$A_{\pi \pi}^{+-} = - C_{\pi \pi}^{+-}$. 

The BABAR~\cite{Aubert:2002jb,Jawahery:2003} and 
BELLE~\cite{Abe:2003ja} measurements were summarized 
last summer at the Lepton-Photon 2003 conference, 
yielding the world averages~\cite{Jawahery:2003}: 
$S_{\pi \pi}^{+-} = -0.58 \pm 0.20$ and 
$C_{\pi \pi}^{+-} = -0.38 \pm 0.16$. 
However, a significant disagreement between the two measurements 
existed and the confidence level that the two are compatible
with each other, in particular in the measurement 
of~$S_{\pi \pi}^{+-}$, was rather low (4.7\%~C.L.). Recently, the 
BELLE collaboration have updated their results for~$S_{\pi \pi}^{+-}$ 
and $A_{\pi \pi}^{+-}$ by including more data. The current BELLE 
measurements~\cite{Abe:2004us} (based on 140~fb$^{-1}$ data) together 
with the updated BABAR results~\cite{Jawahery:2003} (based on 
113~fb$^{-1}$ data) of these quantities are as follows:
\begin{equation}
S_{\pi \pi}^{+-} = \left\{
\begin{array}{ll}
-0.40 \pm 0.22 \pm 0.03  & ({\rm BABAR}) \\
-1.00 \pm 0.21 \pm 0.07  & ({\rm BELLE})   
\end{array}
\right. ,  
\quad
C_{\pi \pi}^{+-} = \left\{
\begin{array}{ll}
-0.19 \pm 0.19 \pm 0.05  & ({\rm BABAR}) \\
-0.58 \pm 0.15 \pm 0.07  & ({\rm BELLE})   
\end{array}
\right. .  
\label{eq:Spipi-Cpipi-new}
\end{equation}
They have been averaged by the Heavy Flavor Averaging 
Group [HFAG] to yield~\cite{HFAG:2004}
\begin{equation}
S_{\pi \pi}^{+-} = -0.74 \pm 0.16 ,
\qquad
C_{\pi \pi}^{+-} = -0.46 \pm 0.13 ,
\label{eq:Spipi-Cpipi-av-new}
\end{equation}
and correspond to~4.6 and~3.5 standard deviation measurements 
from null results,
respectively. It is also reassuring to note that the BELLE and 
BABAR measurements are now closer to each other than was the 
case at the Lepton-Photon 2003 conference, having now a scale
factor~\footnote{We thank the Heavy Flavor Averaging Group and, 
in particular, Andreas H\"ocker, for providing us the updated 
averages and the scale factors.} of~1.7 in $S_{\pi \pi}^{+-}$ 
and~1.4 in $C_{\pi \pi}^{+-}$. Significant updates of the BABAR 
and BELLE results in the $B \to \pi \pi$ decays are awaited later 
this year which will further firm up these measurements.

As can be judged from the results in~(\ref{eq:Spipi-Cpipi-av-new}), 
current measurements of~$S_{\pi \pi}^{+-}$ and~$C_{\pi\pi}^{+-}$ 
have already reached a significant level and invite a theoretical
analysis leading to a determination of the unitarity triangle 
angles~$\alpha$ and hence also~$\gamma$. The importance of these
measurements for the CKM phenomenology has been long anticipated
and discussed at great length in the literature
\cite{Fleischer:2000un,Gronau:2002qj,Fleischer:2002zv,Fleischer:2003xx,%
Buras:2003dj,Buras:2004ub,Lavoura:2004rs}.
Our analysis taking into account the updated $B \to \pi \pi$ data 
has many features which it shares conceptually with the cited 
literature and we shall compare our results with the ones obtained 
in the more recent works~\cite{Buras:2003dj,Buras:2004ub}.
A prerequisite to carry out such an analysis is to get model-independent 
bounds on the non-perturbative dynamical quantities~$|P_c/T_c|$ and 
$\delta_c = \delta_c^P - \delta_c^T$, involving the so-called 
QCD-penguin~$P_c$ and color-allowed tree~$T_c$ topologies.
Here, the subscripts denote that we are using the $c$-convention of 
Gronau and Rosner~\cite{Gronau:2002gj} in choosing the independent 
CKM factors in the analysis of the $B \to \pi \pi$ decays. Discussions 
of the ambiguities in the penguin amplitudes have also been presented
earlier~\cite{Buras:1994pb,London:1999iv,Branco:1999fs}.
Our approach makes use of the isospin-based bounds on the 
ratio~$|P_c/T_c|$ and~$\delta_c$ in the analysis of the data in the 
$B \to \pi \pi$ sector and we show how to incorporate these bounds 
in the analysis of the unitarity triangle in the SM.

There are essentially three parameters~$|P_c/T_c|$, $\delta_c$ 
and~$\alpha$ [the weak phase~$\beta$ is already well measured, see 
Eq.~(\ref{eq:sin2beta-exp})], which can not be determined from the
measurements of just two quantities~$S_{\pi \pi}^{+-}$ 
and~$C_{\pi \pi}^{+-}$. However, correlations and bounds on these
parameters can be obtained which have been presented by the BELLE
collaboration based on their data~\cite{Abe:2003ja,Abe:2004us}.
In the first part of our paper we undertake a similar analysis of 
the combined BABAR and BELLE data and work out the best-fit values 
and  bounds on the parameters~$\delta_c$ and~$|P_c/T_c|$. As our 
analysis is performed within the~SM, we allow the phase~$\beta$ to 
vary in the experimental range and restrict the range of~$\alpha$ 
from the indirect unitarity-triangle (UT) analysis, which we have 
taken from the CKM fitter~\cite{Hocker:2001xe} and another recent 
fit of the CKM parameters~\cite{Ali:2003te}. We first show in this 
paper that the current data on~$S_{\pi \pi}^{+-}$ and~$C_{\pi \pi}^{+-}$ 
restricts the two strong interaction parameters~$\delta_c$ 
and~$|P_c/T_c|$. This information is already helpful in providing 
some discrimination on various competing approaches incorporating 
QCD dynamics in these decays. Conversely, restricting the allowed 
range of~$|P_c/T_c|$ from the current dynamical models, data 
on~$S_{\pi \pi}^{+-}$  and~$C_{\pi \pi}^{+-}$ allows to put 
constraints on~$\alpha$. This has been done by the BELLE 
collaboration~\cite{Abe:2004us}, yielding at 95.5\% C.L.
$90^\circ \leq \phi_2 \leq 146^\circ$ for
$0.15 \leq \vert P_c/T_c \vert \leq 0.45$ and 
$\sin (2\phi_1) = 0.746$. 
However, due to the restrictions on~$|P_c/T_c|$, this remains a 
model-dependent enterprise.

\begin{figure}[tb]
\centerline{
\psfig{width=0.45\textwidth,file=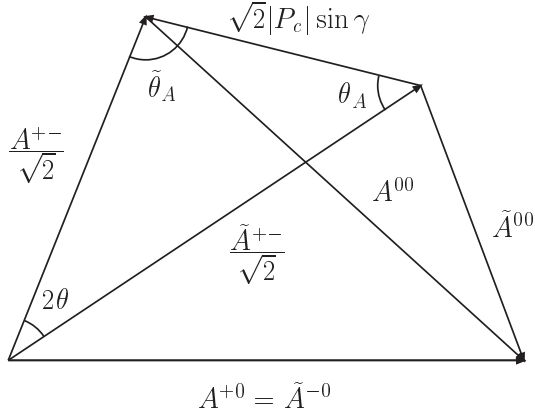}
}
\caption{The isospin triangle for the $B \to \pi \pi$ decay 
         amplitudes~$A^{ij}$ and the same for the phase-shifted 
         charge-conjugate ones 
         $\tilde A^{ij} = {\rm e}^{2 i \gamma} \, \bar A^{ij}$ 
         in the complex plane.}    
\label{fig:isospin-triangle}
\end{figure}

Our analysis differs in this respect from the one carried out by 
the BELLE collaboration. Instead of restricting~$|P_c/T_c|$ by a 
survey of models, we use the isospin symmetry to restrict the range 
of~$|P_c/T_c|$ and~$\delta_c$. To do this, we harness all the current 
data available on the branching ratios for $B^0_d \to \pi^+\pi^-$, 
$B^+ \to \pi^+ \pi^0$ and $B^0_d \to \pi^0 \pi^0$ decays (and their 
charge conjugates), $S_{\pi \pi}^{+-}$ and~$C_{\pi \pi}^{+-}$, and 
study a number of correlations, in particular~$\vert P_c/T_c \vert$ 
vs.~$\cos (2 \theta)$, where~$\theta$ is a penguin-related angle 
which is connected with the relative phase between the amplitudes~$A^{+-}$ 
and~$\tilde A^{+-}$ (see Fig.~\ref{fig:isospin-triangle}). It is 
well known that the isospin symmetry can be used to put a lower bound 
on $\cos (2\theta)$, as first pointed out by Grossman and 
Quinn~\cite{Grossman:1997jr}. Subsequently, the Grossman-Quinn bound was 
improved by Charles~\cite{Charles:1998qx}, who derived in addition a new 
bound involving the $B^0_d \to \pi^0 \pi^0$ and $B^0_d \to \pi^+ \pi^-$ 
decay modes. Based on the observation that the $B \to \pi \pi$ amplitudes
can be represented, using the isospin symmetry, as two (closed) triangles
which have a common base, Gronau et.~al~\cite{Gronau:2001ff} derived an 
improved lower bound on $\cos (2\theta)$~-- the Gronau-London-Sinha-Sinha 
(GLSS) bound. We illustrate this bound numerically using current data 
and the constraints that it implies in the $|P_c/T_c| - \cos (2\theta)$ 
plane for both the $\theta > 0$ and $\theta < 0$ cases, varying~$\gamma$ 
in a large range $25^\circ \leq \gamma \leq 75^\circ$, which adequately 
covers the present range of this angle allowed by the UT fits at 95\%~C.L. 
Lest it be misunderstood, we emphasize that our final results for the
CKM parameters and the dynamical quantities make no restrictions on the
range of $\gamma$, whose value will be returned together with those of the
other quantities by our CKM unitarity fits.
The isospin-based lower bound on $\cos (2\theta)$, and hence an upper 
bound on~$|\theta|$, is a model-independent constraint on the penguin 
contribution in the analysis of the data involving 
the measurements of~$S_{\pi \pi}^{+-}$ and~$C_{\pi \pi}^{+-}$.

There are yet other bounds based on the isospin symmetry in 
$B \to \pi \pi$ decays which lead to restrictions on~$\gamma$. 
In particular, the Buchalla-Safir bound~\cite{Buchalla:2003jr} 
on~$\gamma$ (and its various reincarnations discussed recently 
in the literature~\cite{Botella:2003xp,Lavoura:2004rs}) result 
from the correlations involving $\sin (2\beta)$, $\gamma$, 
$S_{\pi\pi}^{+-}$, and~$C_{\pi\pi}^{+-}$. We have analyzed these 
bounds, but we find that they are not very useful at 
present as  the current central values of $\sin (2\beta)$
and~$- S_{\pi\pi}^{+-}$ almost coincide. For these bounds to be 
useful phenomenologically, the value of $- S_{\pi\pi}^{+-}$ has 
to come down substantially.

In the last part of our analysis, we study the impact of 
the~$S_{\pi \pi}^{+-}$ and~$C_{\pi \pi}^{+-}$ measurements on the 
profile of the unitarity triangle in a model-independent way.
We first show that the quality of the UT fits is not 
modified by the inclusion of the data on~$S_{\pi \pi}^{+-}$ 
and~$C_{\pi \pi}^{+-}$, as the two additional parameters~$|P_c/T_c|$ 
and~$\delta_c$, when  varied in large regions, can always 
reproduce the central values of the~$S_{\pi \pi}^{+-}$ 
and~$C_{\pi \pi}^{+-}$ averages. We then implement the lower bound 
on $\cos (2 \theta)$ in performing the fits of the unitarity triangle
in the $\bar \rho - \bar \eta$ plane. The present bound 
$\cos (2\theta) > 0.27$ removes a small part of the otherwise allowed 
region of the unitarity triangle, but this constraint will become  
more significant in future as the errors on the $B \to \pi \pi$
branching ratios, $S_{\pi \pi}^{+-}$ and~$C_{\pi \pi}^{+-}$ are 
reduced. The effects of the bound on $\cos (2\theta)$ are also shown 
on the correlations $\alpha - \gamma$ and $\cos (2\alpha) - \cos (2\beta)$. 
Working out the $\chi^2$-distributions in the quantities~$\alpha$, 
$|P_c/T_c|$ and~$\delta_c$, we find that the current data prefers  
rather large values for the latter two  quantities, with the minimum
of the $\chi^2$-distributions being at $\vert P_c/T_c \vert = 0.77$ 
and $\delta_c = -43^\circ$. The corresponding best-fit values  
of~$\alpha$ and~$\gamma$ are $\alpha = 92^\circ$ and $\gamma = 64^\circ$. 
At 68\%~C.L., the ranges are:
$81^\circ \leq \alpha \leq 103^\circ$, 
$21.9^\circ \leq \beta \leq 25.5^\circ$,
$54^\circ \leq \gamma \leq 75^\circ$, 
$0.43 \leq \vert P_c/T_c\vert \leq 1.35$, and   
$-64^\circ \leq \delta_c \leq -29^\circ$. 
The bound on $\cos (2\theta)$ is very efficient in 
the exclusion of the large values of~$|P_c/T_c|$ and~$-\delta_c$. 
Their best-fit values are quite a bit larger than anticipated in 
most dynamical approaches. This feature has also been noted in earlier 
studies on the $B \to \pi \pi$ 
decays~\cite{Fleischer:2000un,Buras:2003dj,Buras:2004ub}.
%
%
%

This paper is organized as follows: 
In section~2, we give the relations among the observables in the 
$B^0_d \to \pi^+ \pi^-$ decay and its charge conjugate, the CKM 
parameters and various dynamical quantities. Section~3 contains 
a review of several isospin-based bounds in the $B \to \pi \pi$ 
decays. In section~4, we report on the results of our numerical 
analysis of the time-dependent CP asymmetry in the 
$B_d^0/\bar{B}_d^0 \to \pi^+\pi^-$ decays, and in section~5 we show 
the results of the unitarity triangle fits, correlations involving 
the angles~$\alpha$, $\beta$ and~$\gamma$, and the dynamical 
quantities~$\vert P_c/T_c\vert$ and~$\delta_c$, carried out in the
context of the~SM. We conclude with a summary and some 
remarks in section~6.

\section{Relations among the Observables in the $B \to \pi \pi$ Decays,
         CKM Parameters and Dynamical Quantities}

In this section we present the analytic formulae that we need 
to discuss the time-dependent CP asymmetry and the branching 
ratio in the $B^0_d \to \pi^+ \pi^-$ decay, and their relations 
with the CKM parameters and various dynamical quantities.
 
The amplitudes $A^{+ -} \equiv A [B^0_d \to \pi^+ \pi^-]$ 
and its charge-conjugate $\bar A^{+ -} \equiv A [\bar B^0_d 
\to \pi^+ \pi^-]$ can be written by using the Gronau-Rosner 
$c$-convention~\cite{Gronau:2002gj} as follows: 
\begin{eqnarray} 
A^{+ -} & = & 
V_{ub}^* V_{ud} \, A^{+ -}_u + V_{cb}^* V_{cd} \, A^{+ -}_c + 
V_{tb}^* V_{td} \, A^{+ -}_t 
\label{eq:A-def} \\ 
& = & V_{ub}^* V_{ud} \left ( A^{+ -}_u - A^{+ -}_t \right ) + 
V_{cb}^* V_{cd} \left ( A^{+ -}_c - A^{+ -}_t \right ) 
\nonumber \\ 
& \equiv & - \left ( 
|T_c| \, e^{i \delta^T_c} \, e^{+ i \gamma} + 
|P_c| \, e^{i \delta^P_c} 
\right ) , 
\nonumber \\ 
\bar A^{+ -} & = & 
V_{ub} V_{ud}^* \left ( A^{+ -}_u - A^{+ -}_t \right ) +
V_{cb} V_{cd}^* \left ( A^{+ -}_c - A^{+ -}_t \right )
\label{eq:A-bar-def} \\ 
& \equiv & - \left (
|T_c| \, e^{i \delta^T_c} \, e^{- i \gamma} + 
|P_c| \, e^{i \delta^P_c}
\right ) . 
\nonumber 
\end{eqnarray}
In getting the last expressions for the amplitudes the unitarity 
relation~(\ref{eq:vudtriangle}) 
has been used together with the phase convention 
$V_{ub} = |V_{ub}| \, e^{- i \gamma}$ for this CKM matrix element. 

The phenomenon of the $B^0_d - \bar B^0_d$ mixing modulates the 
time dependence of the decay amplitudes for $B^0_d (t) \to \pi^+\pi^-$ 
and $\bar B^0_d (t) \to \pi^+ \pi^-$:
\begin{eqnarray}
{\cal A}^{+ -} (t) & = & e^{- i M_B t} e^{- \Gamma t/2}
\bigg \{ \cos \frac{\Delta M_B t}{2} \left [ 
\cosh \frac{\Delta \Gamma t}{4} - \lambda_{\pi \pi}^{+ -} 
\sinh \frac{\Delta \Gamma t}{4} \right ] 
\label{eq:A-time} \\
& + & i \sin \frac{\Delta M_B t}{2} \left [ 
\lambda_{\pi \pi}^{+ -} \cosh \frac{\Delta \Gamma t}{4} - 
\sinh \frac{\Delta \Gamma t}{4} \right ]
\bigg \} A^{+ -} ,
\nonumber \\
\bar {\cal A}^{+ -} (t) & = & e^{- i M_B t} e^{- \Gamma t/2}
\bigg \{ \cos \frac{\Delta M_B t}{2} \left [ 
\lambda_{\pi \pi}^{+ -} \cosh \frac{\Delta \Gamma t}{4} - 
\sinh \frac{\Delta \Gamma t}{4} \right ]
\label{eq:A-bar-time} \\ 
& + & i \sin \frac{\Delta M_B t}{2} \left [ 
\cosh \frac{\Delta \Gamma t}{4} - \lambda_{\pi \pi}^{+ -} 
\sinh \frac{\Delta \Gamma t}{4} \right ]
\bigg \} \frac{p}{q} \, A^{+-} .
\nonumber
\end{eqnarray}
Here, $M_B$ and $\Gamma$ are the average mass and decay width
of the $B^0_d - \bar B^0_d$ system, and $\Delta M_B$ and $\Delta \Gamma$ 
are the the mass- and width- difference in the two mass eigenstates,
 respectively, 
$p/q \simeq V_{td}^*/V_{td} = e^{2 i \beta}$ is the mixing parameter,
and the quantity:
\begin{equation}
\lambda_{\pi \pi}^{+ -} = \frac{q}{p} \, \frac{\bar A^{+-}}{A^{+-}}
= e^{2 i \alpha} \,
\frac{1 + |P_c/T_c| \, e^{i \delta_c} \, e^{+ i \gamma}}
     {1 + |P_c/T_c| \, e^{i \delta_c} \, e^{- i \gamma}}
\equiv |\lambda_{\pi \pi}^{+ -}| \, e^{2 i \alpha_{\rm eff}},
\label{eq:lambda-def}
\end{equation}
is introduced which encodes all the information about the 
CP asymmetry in this decay. Here, $\alpha_{\rm eff} = \alpha + \theta$, 
and~$\theta$ is the penguin-pollution parameter shown in 
Fig.~\ref{fig:isospin-triangle}, which is connected  with the 
relative phase between the amplitudes~$A^{+-}$ and~$\bar A^{+-}$, 
$\Delta \phi^{+-} = 2 (\gamma + \theta)$, 
and the relation $\alpha + \beta + \gamma = \pi$ has been used. 
Note that in the limit $P_c/T_c \to 0$, $\theta \to 0$ and 
$\alpha_{\rm eff} \to \alpha$.

The partial decay widths of the time-dependent 
$B^0_d (t) \to \pi^+ \pi^-$ and $\bar B^0_d (t) \to \pi^+ \pi^-$ 
decays are proportional, respectively, to~\cite{Quinnsanda:2002}
\begin{eqnarray}
|{\cal A}^{+ -} (t)|^2 & = & 
e^{- \Gamma t} \, B_{\pi \pi}^{+ -}
\left [ 1 + C_{\pi \pi}^{+ -} \cos (\Delta M_B t)  
- S_{\pi \pi}^{+ -} \sin (\Delta M_B t) \right ] , 
\label{eq:A-time-squared} \\
|\bar {\cal A}^{+ -} (t)|^2 & = & 
e^{- \Gamma t} \, B_{\pi \pi}^{+ -} 
\left [ 1 - C_{\pi \pi}^{+ -} \cos (\Delta M_B t) 
+ S_{\pi \pi}^{+ -} \sin (\Delta M_B t) \right ] , 
\label{eq:A-bar-time-squared}
\end{eqnarray}
where $|p/q| = 1$ is used and the following quantities are
introduced:  
\begin{eqnarray} 
B_{\pi \pi}^{+ -} & = & 
\frac{1}{2} \left [ |A^{+ -}|^2 + |\bar A^{+ -}|^2 \right ] = 
\frac{1}{2} \left [ 1 + |\lambda_{\pi \pi}^{+ -}|^2 \right ] 
|A^{+ -}|^2 , 
\label{eq:Bpp-pm-def} \\ 
C_{\pi \pi}^{+ -} & = &  
\frac{|A^{+ -}|^2 - |\bar A^{+ -}|^2}
     {|A^{+ -}|^2 + |\bar A^{+ -}|^2} = 
\frac{1 - |\lambda_{\pi \pi}^{+ -}|^2}
     {1 + |\lambda_{\pi \pi}^{+ -}|^2} ,
\label{eq:Cpp-pm-def} \\ 
S_{\pi \pi}^{+ -} & = & 
\frac{2 \, {\rm Im} \left [ (q/p) \bar A^{+ -} (A^{+ -})^* \right ]} 
     {|A^{+ -}|^2 + |\bar A^{+ -}|^2} =
\frac{2 \, {\rm Im} \, \lambda_{\pi \pi}^{+ -}}
     {1 + |\lambda_{\pi \pi}^{+ -}|^2} \equiv 
y_{\pi \pi}^{+ -} \, \sin (2 \alpha_{\rm eff}), 
\label{eq:Spp-pm-def} \\ 
y_{\pi \pi}^{+ -} & = & 
\frac{2 \, |A^{+ -}| \, |\bar A^{+ -}|} 
     {|A^{+ -}|^2 + |\bar A^{+ -}|^2} =
\frac{2 \, |\lambda_{\pi \pi}^{+ -}|}
     {1 + |\lambda_{\pi \pi}^{+ -}|^2} .  
\label{eq:ypp-pm-def}  
\end{eqnarray}
Using the expression  for $\lambda_{\pi \pi}^{+-}$~ given in 
Eq.~(\ref{eq:lambda-def}), the above 
quantities can be rewritten in the following form: 
\begin{eqnarray}
B_{\pi \pi}^{+-} & \equiv & |T_c|^2 \, R_{\pi \pi}^{+-} = 
|T_c|^2 + 2 |P_c| |T_c| \cos \delta_c \cos \gamma + |P_c|^2 , 
\label{eq:B-pipi-pm-angles} \\  
C_{\pi \pi}^{+-} & = & \frac{2}{R_{\pi \pi}^{+-}} \,  
\left | \frac{P_c}{T_c} \right | \sin \delta_c \sin \gamma , 
\label{eq:C-pipi--pm-angles} \\
S_{\pi \pi}^{+-} & = & \frac{1}{R_{\pi \pi}^{+-}} \left [ 
\sin (2 \alpha) - 2 \left | \frac{P_c}{T_c} \right | \cos \delta_c 
\sin (\alpha - \beta) - \left | \frac{P_c}{T_c} \right |^2 
\sin (2 \beta) \right ] , 
\label{eq:S-pipi-pm-angles} \\ 
y_{\pi \pi}^{+-} & = & \sqrt{1 - \left ( C_{\pi \pi}^{+-} \right )^2}. 
\label{eq:y-pipi-pm-angles} 
\end{eqnarray}
Making the back transformation, $|T_c|$, $|P_c|$ and $\delta_c$ 
can be expressed as follows: 
\begin{eqnarray}
|T_c|^2 & = & \frac{B_{\pi \pi}^{+-}}{1 - \cos (2\gamma)}
\left [ 1 - y_{\pi \pi}^{+-} \cos (2 \theta - 2 \gamma) \right ] ,
\label{eq:Tc2} \\
|P_c|^2 & = & \frac{B_{\pi \pi}^{+-}}{1 - \cos (2\gamma)}
\left [ 1 - y_{\pi \pi}^{+-} \cos (2 \theta) \right ] ,
\label{eq:Pc2} \\
\tan \delta_c & = &
\frac{C_{\pi \pi}^{+-} \sin \gamma}
     {y_{\pi \pi}^{+-} \cos (2 \theta - \gamma) - \cos \gamma} .
\label{eq:tan-delta}
\end{eqnarray}
In the limit of neglecting the penguin 
contribution (i.e., $|P_c| \to 0$), the CP-asymmetry coefficient 
$C_{\pi \pi}^{+-}$ goes to zero (and $y_{\pi \pi}^{+-} \to 1$) 
as well as~$\theta \to 0$, in agreement with Eq.~(\ref{eq:Pc2}). 
Also, in this limit, $B_{\pi \pi}^{+-} = |T_c|^2$, as in this case 
the branching ratio is completely defined by the tree contribution.

\begin{figure}[tbp]
\centerline{
\psfig{width=0.45\textwidth,file=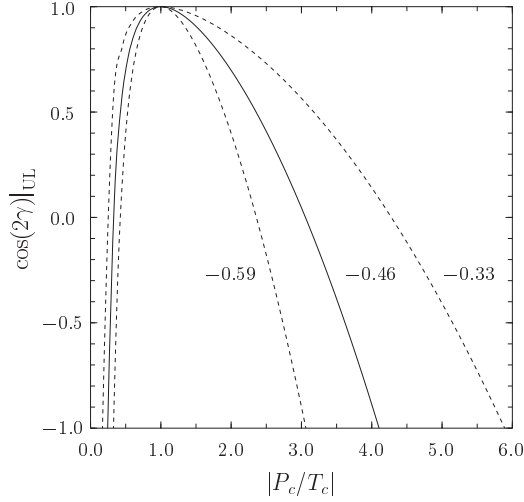}
}
\caption{The dependence of the upper limit $\cos(2\gamma)_{\rm UL}$ 
         on the ratio~$|P_c/T_c|$ for 
         $C_{\pi \pi}^{+-} = -0.46$ (the current central value), 
         $C_{\pi \pi}^{+-} = -0.59$ and $C_{\pi \pi}^{+-} = -0.33$,
         which demarcate the $\pm 1 \sigma$ experimental measurements.}
\label{fig:c2gUp-PT}
\end{figure}

In terms of~$\gamma$ and~$\theta$, the penguin-to-tree ratio 
squared has the following expression:
\begin{equation}
r_c^2 \equiv \left | \frac{P_c}{T_c} \right|^2 =
\frac{1 - y_{\pi \pi}^{+-} \cos(2 \theta)}
     {1 - y_{\pi \pi}^{+-} \cos(2 \theta - 2 \gamma)} 
= \frac{1 - y_{\pi \pi}^{+-} \cos(2 \alpha_{\rm eff} - 2 \alpha)}
     {1 - y_{\pi \pi}^{+-} \cos(2 \alpha_{\rm eff} + 2 \beta)}.
\label{eq:PcTc2}
\end{equation}
This relation constrains~$r_c$ in terms of $\cos (2\theta)$, 
given~$\gamma$ and~$y_{\pi\pi}^{+-}$. It should be noted that 
for fixed values of~$y_{\pi\pi}^{+-}$ and~$r_c$, $\cos (2\gamma)$ 
varies in the range:
\begin{equation}
-1 \le \cos (2\gamma) \le
\frac{1 - (C_{\pi\pi}^{+-})^2 (1 + r_c^4)/(2 r_c^2)}
     {1 - (C_{\pi\pi}^{+-})^2}.
\end{equation}
It is easy to see that the upper limit of $\cos(2\gamma)$ is 
equal to~1 when $r_c = 1$, independent of~$C_{\pi\pi}^{+-}$. 
For $r_c \neq 1$, the allowed range of $\cos (2\gamma)$ puts 
a constraint in the $r_c - C_{\pi\pi}^{+-}$ plane. Thus, the 
allowed domain of~$r_c$ is completely
defined by the direct CP-asymmetry coefficient and, for
negative~$C_{\pi\pi}^{+-}$ (in accordance with the experimental
data), it is given by
\begin{equation}
(r_c)_{\rm min, max} = - 
\frac{1 \pm \sqrt{1 - (C_{\pi\pi}^{+-})^2}}{C_{\pi\pi}^{+-}} .
\end{equation}
In particular, for the central experimental value
$C_{\pi\pi}^{+-} = -0.46$, the allowed range of~$r_c$ is as follows:
\begin{equation}
0.244 \le r_c \le 4.104 .
\end{equation}
For the current experimental central value of~$C_{\pi \pi}^{+-}$, 
and its $\pm 1 \sigma$ limits, the dependence of the  upper limit
 $\cos(2\gamma)|_{\rm UL}$
 on the magnitude of the penguin-to-tree ratio 
is presented in Fig.~\ref{fig:c2gUp-PT}. Decreasing the magnitude 
of~$C_{\pi\pi}^{+-}$, the allowed region of~$r_c$ becomes wider.
%
%
%
%
The expression for $\cos (2\theta)$ as a function
of~$y_{\pi\pi}^{+-}$, $r_c$ and $\cos(2\gamma)$ is as follows:
\begin{equation}
\cos (2 \theta) =
\frac{(1 - r_c^2) [1 - r_c^2 \cos(2\gamma)] \pm r_c^2
\sqrt{[1 - \cos^2 (2\gamma)] 
      \{ 2 (y_{\pi\pi}^{+-})^2 r_c^2 [1 - \cos (2\gamma)] - 
         (1 - r_c^2)^2 [1 - (y_{\pi\pi}^{+-})^2] \}}}
  {y_{\pi\pi}^{+-} \{(1 - r_c^2)^2 + 2 r_c^2 [1 - \cos (2\gamma)] \}} .
\end{equation}

\section{Isospin-Based Bounds in $B \to \pi\pi$ Decays}

With partial experimental information on $B \to \pi \pi$ decays 
available at present, it is of practical importance to 
get useful restrictions on the dynamical parameters   
at hand, $r_c$ and $\delta_c$. It is obvious from~(\ref{eq:tan-delta}) 
and~(\ref{eq:PcTc2}), that apart from $C_{\pi\pi}^{+-}$, which 
is measured from the time-dependent CP asymmetry, the angle~$2\theta$ 
plays a central role in constraining the dynamical parameters 
of interest. While $2 \theta$ will be determined eventually from 
the measurement of~$A^{00}$ and~$\tilde{A}^{00}$ 
(see Fig.~\ref{fig:isospin-triangle}), this is not the case now. 
Instead, several bounds have been derived on $\cos (2\theta)$ 
using the isospin symmetry, which we review here. The first of these 
which we will work out numerically is the GLSS bound~\cite{Gronau:2001ff}: 
\begin{equation} 
\cos (2\theta) \ge 
\frac{(B_{\pi\pi}^{+-} + 2 B_{\pi\pi}^{+0} - 2 B_{\pi\pi}^{00})^2 
       - 4 B_{\pi\pi}^{+-} B_{\pi\pi}^{+0}}
     {4 y_{\pi\pi}^{+-} B_{\pi\pi}^{+-} B_{\pi\pi}^{+0}} ,  
\label{eq:GLSS-bound}
\end{equation}
where $B_{\pi\pi}^{+0}$ and $B_{\pi\pi}^{00}$ are the quantities 
constructed from the $B^+ \to \pi^+ \pi^0$ and $B^0_d \to \pi^0 \pi^0$ 
decay amplitudes in a similar way as in Eq.~(\ref{eq:Bpp-pm-def}).  
It was also demonstrated by Gronau et.~al~\cite{Gronau:2001ff} 
that this bound is stronger than  both the 
Grossman-Quinn and Charles bounds. As a byproduct, a 
bound on the direct CP asymmetry $C_{\pi\pi}^{00}$ in the 
$B^0_d \to \pi^0 \pi^0$ decay was also obtained by these
 authors~\cite{Gronau:2001ff}: 
\begin{equation}
C_{\pi\pi}^{00} \ge C_{\pi\pi}^{+-} \, 
\frac{B_{\pi\pi}^{+-} \, 
      (B_{\pi\pi}^{+-} - 2 B_{\pi\pi}^{+0} - 2 B_{\pi\pi}^{00})} 
     {2 B_{\pi\pi}^{00} \, 
      (B_{\pi\pi}^{+-} + 2 B_{\pi\pi}^{+0} - 2 B_{\pi\pi}^{00})} . 
\label{eq:Cpipi00-GLSS-bound}
\end{equation}

There have also been attempts~\cite{Giri:2002aw} to derive isospin 
bounds on $\cos(2\theta)$ in the $B^0_d \to \pi^+ \pi^-$ 
decay which are based on the knowledge of the direct CP asymmetry 
$C_{\pi\pi}^{+-}$ alone. As for the GLSS bound~\cite{Gronau:2001ff}, 
the starting point is the isospin-based triangular relation between 
the~$A^{+-}$, $A^{00}$ and~$A^{+0}$ amplitudes: 
\begin{equation} 
\frac{A^{+-}}{\sqrt 2} + A^{00} = A^{+0}, 
\label{eq:isospin-triangle}
\end{equation}
corresponding to the $B^0_d \to \pi^+ \pi^-$, 
$B^0_d \to \pi^0 \pi^0$ and $B^+ \to \pi^+ \pi^0$ decays,   
respectively, and a similar one for the phase-shifted
charged-conjugate amplitudes 
$\tilde A^{ij} = {\rm e}^{2 i \gamma} \, \bar A^{ij}$. 
 The graphical representation of 
both triangles is shown in Fig.~\ref{fig:isospin-triangle}.  
It should be noted that the magnitude of the difference 
between the~$A^{+-}$ and~$\tilde A^{+-}$ amplitudes is 
$\sqrt 2 |P_c| \sin \gamma$ and not $\sqrt 2 |P_t| \sin \alpha$ 
as the $c$-convention~\cite{Gronau:2002gj} is employed 
for the amplitudes throughout this paper. With the help of the 
sine theorem, $\sin |2\theta|$ can be written as:  
\begin{equation}
\sin |2\theta| = \frac{2 |P_c| \sin \gamma}{|A^{+-}|} \sin \theta_A 
= \frac{2 |P_c| \sin \gamma}{|\bar A^{+-}|} \sin \tilde \theta_A . 
\label{eq:sin2theta}
\end{equation}
Squaring all the terms, the above relation can be rewritten 
as two inequalities: 
\begin{equation}
\sin^2 |2\theta| \le 2 \, 
\frac{1 - y_{\pi \pi}^{+-} \cos (2\theta)}
     {1 + C_{\pi \pi}^{+-}} , 
\qquad 
\sin^2 |2\theta| \le 2 \, 
\frac{1 - y_{\pi \pi}^{+-} \cos (2\theta)}
     {1 - C_{\pi \pi}^{+-}} , 
\label{eq:sin2theta-inequalities}
\end{equation}
following from the conditions $\sin^2 \theta_A \leq 1$ and 
$\sin^2 \tilde{\theta}_A \leq 1$, respectively. Here, 
Eqs.~(\ref{eq:Bpp-pm-def}), (\ref{eq:Cpp-pm-def}) and~(\ref{eq:Pc2}) 
were used to eliminate~$|A^{+-}|^2$, $|\bar A^{+-}|^2$ and~$|P_c|^2$.
While $\sin^2 |2\theta| \leq 1 $, this does not imply that the 
expressions on the r.h.s. of~(\ref{eq:sin2theta-inequalities}) also
satisfy this upper bound. Hence, no bound on~$|2\theta|$ follows 
from Eq.~(\ref{eq:sin2theta-inequalities})\footnote{
We are grateful to David London and Nita and Rahul Sinha for pointing 
this out to us.} and the GLSS bounds are indeed the strongest 
isospin-based bounds in the $B \to \pi \pi$ sector.

In addition to the above bounds on $\cos (2\theta)$, bounds on the CKM 
angle~$\gamma$ have also been derived in the literature recently which 
are based on the study of the correlation $\gamma - S_{\pi\pi}^{+-}$, 
given $\sin (2\beta)$~\cite{Buchalla:2003jr,Botella:2003xp,Lavoura:2004rs}. 
We reproduce these bounds below and discuss their impact in the 
next section. Relating the unitarity triangle angles~$\beta$ and~$\gamma$ 
with the Wolfenstein parameters~$\bar \rho$ and~$\bar \eta$: 
\begin{equation} 
1 - \bar \rho \pm i \bar \eta = R_t \, e^{\pm i \beta} , 
\qquad 
\bar \rho \pm i \bar \eta = 
R_b \, e^{\pm i \gamma} , 
\label{eq:CKM-angles}
\end{equation}
where $R_t$ and~$R_b$ are defined in Eq.~(\ref{eq:Rb-Rt}), 
the quantities~$R_{\pi \pi}^{+-}$, $C_{\pi \pi}^{+-}$ and 
$S_{\pi \pi}^{+-}$ can be expressed in the form: 
\begin{eqnarray} 
R_{\pi \pi}^{+-} & = & 
1 + \frac{2 \bar \rho}{R_b} 
\left | \frac{P_c}{T_c} \right | \cos \delta_c + 
\left | \frac{P_c}{T_c} \right |^2 ,  
\label{Rpp-pm-rho-eta} \\ 
C_{\pi \pi}^{+-} & = & 
\frac{2 \bar \eta}{R_b R_{\pi \pi}^{+-}} 
\left | \frac{P_c}{T_c} \right | \sin \delta_c ,  
\label{eq:Cpp-pm-rho-eta} \\ 
S_{\pi \pi}^{+-} & = &  
\frac{- 2 \bar \eta}{R_b^2 R_t^2 R_{\pi \pi}^{+-}} \, 
\left [ \bar \rho - R_b^2 + (1 - R_b^2) R_b  
\left | \frac{P_c}{T_c} \right | \cos \delta_c 
+ (1 - \bar \rho) R_b^2 \left | \frac{P_c}{T_c} \right |^2 
\right ] .   
\label{eq:Spp-pm-rho-eta}
\end{eqnarray} 
The relation for $S_{\pi \pi}^{+-}$ given above agrees with Eq.~(5) 
of the paper by Buchalla and Safir~\cite{Buchalla:2003jr}, if one 
introduces the pure strong-interaction quantity
\begin{equation}
r \equiv R_b \left | \frac{P_c}{T_c} \right |,
\label{eq:r-def}
\end{equation}
used by these authors. Note also that the equation for 
$C_{\pi \pi}^{+-}$ can be rewritten in terms of this 
quantity~$r$ in the following form: 
\begin{equation} 
\left ( \bar \rho + r \cos \delta_c \right )^2 + 
\left ( \bar \eta - \frac{r \sin \delta_c}{C_{\pi \pi}^{+-}} 
\right )^2 = 
\left (  \frac{y_{\pi \pi}^{+-}}{C_{\pi \pi}^{+-}} \, 
r \sin \delta_c \right )^2 . 
\label{eq:C-pipi-circle}  
\end{equation}
For the phenomenological analysis, it is more convenient to 
eliminate~$\bar \rho$ from Eqs.~(\ref{eq:Cpp-pm-rho-eta}) 
and~(\ref{eq:Spp-pm-rho-eta}) with the help of the 
relation~\cite{Buchalla:2003jr}: 
\begin{equation} 
1 - \bar \rho = \bar \eta \cot \beta \equiv \bar \eta \, \tau.   
\label{eq:tau-def}
\end{equation}
With this, the Wolfenstein parameter~$\bar \eta$ can be related 
to either~$S_{\pi \pi}^{+-}$ or~$C_{\pi \pi}^{+-}$ as follows: 
\begin{eqnarray} 
\bar \eta^{(S)} & = & \frac{1}{(1 + \tau^2) S_{\pi \pi}^{+-}} 
\bigg [ (1 + \tau \, S_{\pi \pi}^{+-}) (1 + r \cos \delta_c) 
\label{eq:eta-bar-Spipi} \\ 
& \pm &  
\sqrt{(1 - S_{\pi \pi}^{+-})^2 (1 + r \cos \delta_c)^2 - 
      (1 + \tau^2) S_{\pi \pi}^{+-} 
      [S_{\pi \pi}^{+-} + \sin (2 \beta)] r^2 \sin^2 \delta_c} 
\bigg ] , 
\nonumber \\ 
\bar \eta^{(C)} & = & 
\frac{1}{2} (1 + r \cos \delta_c) \sin (2 \beta) 
+ \frac{1}{(1 + \tau^2) C_{\pi \pi}^{+-}} 
\bigg [ r \sin \delta_c 
\label{eq:eta-bar-Cpipi} \\ 
& \pm &  
\sqrt{(1 + \tau^2) (y_{\pi \pi}^{+-})^2 r^2 \sin^2 \delta_c - 
      \left [C_{\pi \pi}^{+-} (1 + r \cos \delta_c) - 
             \tau \, r \sin \delta_c \right ]^2}   
\bigg ] . 
\nonumber 
\end{eqnarray}
The first of these relations has been obtained by Buchalla and 
Safir (BS)~\cite{Buchalla:2003jr}, and has been used to derive an 
upper bound on~$\bar \eta$ and, hence, a lower bound on~$\gamma$: 
\begin{equation}
\gamma \ge \frac{\pi}{2} - \arctan 
\frac{S_{\pi \pi}^{+-} - \tau 
         \Big [ 1 - \sqrt{1 - (S_{\pi \pi}^{+-})^2} \, \Big ]}
     {1 + \tau S_{\pi \pi}^{+-} - \sqrt{1 - (S_{\pi \pi}^{+-})^2}} , 
\label{eq:gamma-BuchSaf-bound}    
\end{equation}
which holds in the range $-\sin(2\beta) \leq S_{\pi \pi}^{+-} \leq 1$. 
However, the current central experimental 
value~(\ref{eq:Spipi-Cpipi-av-new}) of~$S_{\pi \pi}^{+-}$
practically coincides with the central value~(\ref{eq:sin2beta-exp}) 
of~$-\sin (2\beta)$, and hence no useful bound on the CKM 
angle~$\gamma$ follows from the BS bound at present. We shall show 
this bound as a function of~$S_{\pi \pi}^{+-}$, as well as its 
extension for the case of $C_{\pi \pi}^{+-} \neq 0$: 
\begin{equation} 
\tan \gamma \ge L_- = 
\frac{1 + S_{\pi \pi}^{+-} \sin(2\beta) + 
      \sqrt{1 - (C_{\pi \pi}^{+-})^2 - (S_{\pi \pi}^{+-})^2}
      \cos(2\beta)}
     {\sqrt{1 - (C_{\pi \pi}^{+-})^2 - (S_{\pi \pi}^{+-})^2} 
      \sin(2\beta) - S_{\pi \pi}^{+-} \cos(2\beta)} , 
\label{eq:gamma-BotSil-bound}
\end{equation}
obtained by Botella and Silva~\cite{Botella:2003xp}. The next step 
in the generalization of the BS bound was recently undertaken by 
Lavoura~\cite{Lavoura:2004rs} who considered the modification of 
this bound by putting restrictions on the strong phase~$\delta_c$. 
With the current experimental values of~$S_{\pi\pi}^{+-}$ 
and~$C_{\pi\pi}^{+-}$, also the Botella-Silva and 
Lavoura versions of the BS bound are currently not useful in
constraining~$\gamma$. With precise measurements of~$C_{\pi \pi}^{+-}$ 
and~$S_{\pi \pi}^{+-}$ in future, 
these bounds may, in any case, provide useful consistency checks for 
the dynamical models used in the estimates of~$r$ and~$\delta_c$.

\section{Numerical Analysis of the CP Asymmetry in 
         $B^0_d \to \pi^+ \pi^-$ Decay} 

Within the~SM, the targets for the experiments measuring 
the angles~$\alpha$ and~$\gamma$ are fairly well
defined,  as the fits of the unitarity triangle through the
measurements of the CKM matrix elements yield the following 
ranges for these angles at 95\%~C.L.~\cite{Ali:2003te}:
\begin{equation}
70^\circ \leq \alpha \leq 115^\circ~, 
\qquad  
43^\circ \leq \gamma \leq 86^\circ~.
\label{eq:phi13-ckm-ali}
\end{equation}
The corresponding 95\%~C.L. ranges obtained using the default values 
of the input parameters by the CKM fitter group~\cite{Hocker:2001xe} 
are very similar: 
\begin{equation}
77^\circ \leq \alpha \leq 122^\circ~, 
\qquad 
37^\circ \leq \gamma \leq 80^\circ~.
\label{eq:phi13-ckm-fitter}
\end{equation}
So, if the SM is correct, and currently there is no experimental reason 
to believe otherwise, then from the $B \to \pi \pi$ analysis, values 
of~$\alpha$ and~$\gamma$ should emerge which are compatible with their 
anticipated ranges listed above. Of course, the hope is that direct 
measurements of these angles will greatly reduce the currently allowed 
ranges. However, for this to happen, one has to 
determine the dynamical quantities~$|P_c/T_c|$ and~$\delta_c$.

Surveying the recent literature on the estimates of~$|P_c/T_c|$ 
and~$\delta_c$ in $B \to \pi \pi$ decays, we remark that they are 
either based on specific schemes based on factorization in which 
non-factorizing effects are implemented using perturbative QCD in 
the large-$m_b$ limit~\cite{Beneke:1999br,Keum:2000ph}, 
or on phenomenological approaches based on some input from other data 
and factorization. A typical study in the latter case makes use of 
the data on the $B \to \pi \ell \nu_\ell$ and $B \to K \pi$ decays, 
which are used in conjunction with the assumption of factorization
and estimates of the $SU(3)$-breaking effects~\cite{Luo:2003hn}.
Some representative estimates in these approaches are as follows: 
$\vert P_c/T_c \vert = 0.285 \pm 0.076$
[Beneke, Buchalla, Neubert, Sachrajda]~\cite{Beneke:2001ev},
$\vert P_c/T_c \vert = 0.32 ^{+0.16}_{-0.09}$ 
[Beneke, Neubert]~\cite{Beneke:2003zv},
$\vert P_c/T_c \vert = 0.29 \pm 0.09$ 
[Buchalla, Safir]~\cite{Buchalla:2003jr},
$\vert P_c/T_c \vert = 0.23 ^{+0.07}_{-0.05}$ 
[Keum, Sanda]~\cite{Keum:2002vi},
$\vert P_c/T_c \vert = 0.276 \pm 0.064$ 
[Gronau, Rosner]~\cite{Gronau:1990ka}, 
$\vert P_c/T_c \vert = 0.26 \pm 0.08$ 
[Luo, Rosner]~\cite{Luo:2003hn}.
(See, also Xiao {\it et al.}~\cite{Xiao:2003bc}). 
Thus, $\vert P_c/T_c \vert=0.30$ is a typical value from these estimates. 

What concerns the strong phase difference $\delta_c$, the two dynamical
approaches developed in detail (QCD-Factorization~\cite{Beneke:1999br}
and pQCD~\cite{Keum:2000ph}) differ considerably from each other
due to a different power counting and the treatment of the annihilation
contributions in the decay amplitudes. When comparing the current data 
with these specific approaches, we shall take for the sake of definiteness
the estimates by Buchalla and Safir~\cite{Buchalla:2003jr} to represent 
the QCD-factorization approach, $|P_c/T_c| = 0.29 \pm 0.09$ and 
$\delta_c = 0.15 \pm 0.25$ radians ($\delta_c = 9^\circ \pm 15^\circ$), 
and the estimates by Keum and Sanda~\cite{Keum:2003qi}, 
$|P_c/T_c| = 0.23^{+0.07}_{-0.05}$ and 
$-41^\circ \leq \delta_c \leq -32^\circ$, for the pQCD approach. 
Within the SM, the consistency test of these approaches lies in an 
adequate description of the data on~$S_{\pi \pi}^{+-}$ 
and~$C_{\pi\pi}^{+-}$, with the parameters~$\alpha$, $\gamma$, $|P_c/T_c|$ 
and~$\delta_c$ all lying in their specified ranges. However, as~$|P_c/T_c|$ 
and~$\delta_c$ are not known directly from data or a first principle 
calculation, we can leave them as free parameters and determine them 
from the overall fits. We shall pursue both approaches in this section.

We now present our numerical analysis of the current 
averages of~$S_{\pi \pi}^{+-}$ and~$C_{\pi \pi}^{+-}$ 
given in~(\ref{eq:Spipi-Cpipi-av-new}).  
For the construction of the~C.L. contours, the following 
$\chi^2$-function is used: 
\begin{equation} 
\chi^2 = \left [ 
\frac{C_{\pi \pi}^{+-} - (C_{\pi \pi}^{+-})_{\rm exp}}
     {\Delta C_{\pi \pi}^{+-}} \right ]^2 + 
\left [ \frac{S_{\pi \pi}^{+-} - (S_{\pi \pi}^{+-})_{\rm exp}}
     {\Delta S_{\pi \pi}^{+-}} \right ]^2 , 
\label{eq:chi2}
\end{equation}
which is equated to 2.30, 6.18, and 11.83, corresponding to 68.3\%, 
95.5\%, and 99.7\%~C.L., respectively, for two degrees of freedom.  

\begin{figure}[tbp!]
\centerline{
\psfig{width=0.40\textwidth,file=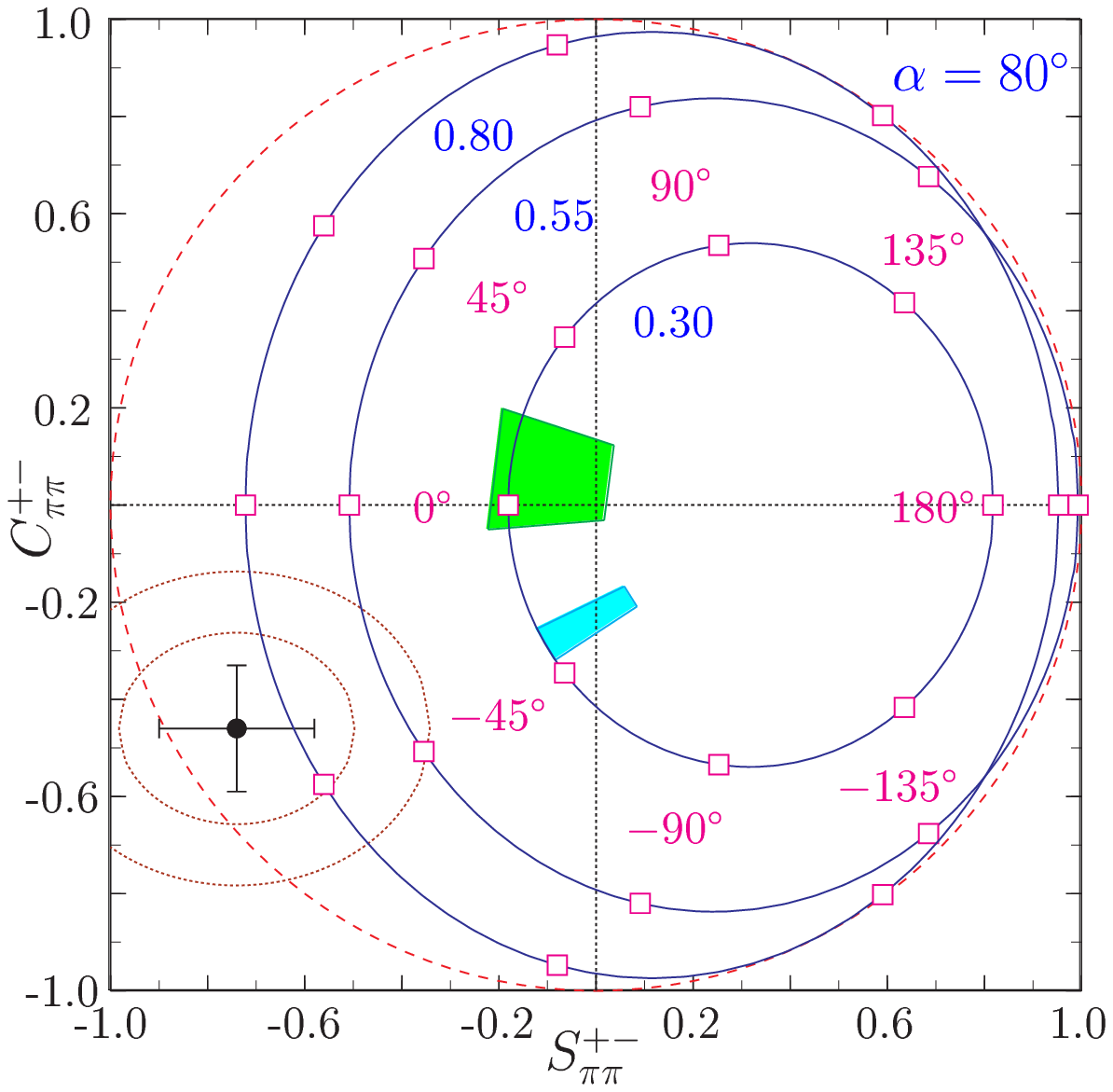}
\quad
\psfig{width=0.40\textwidth,file=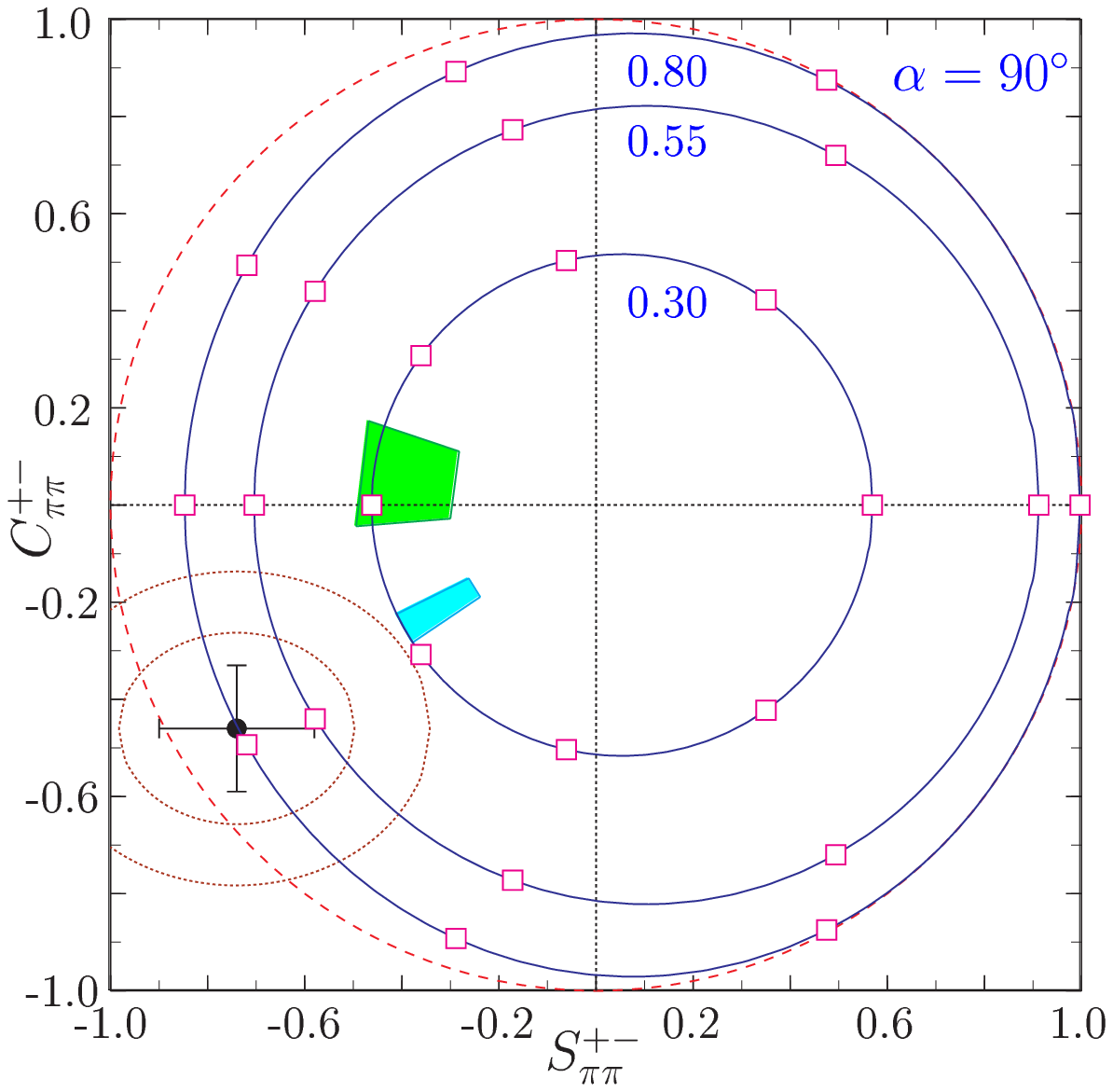}
}
\bigskip
\centerline{
\psfig{width=0.40\textwidth,file=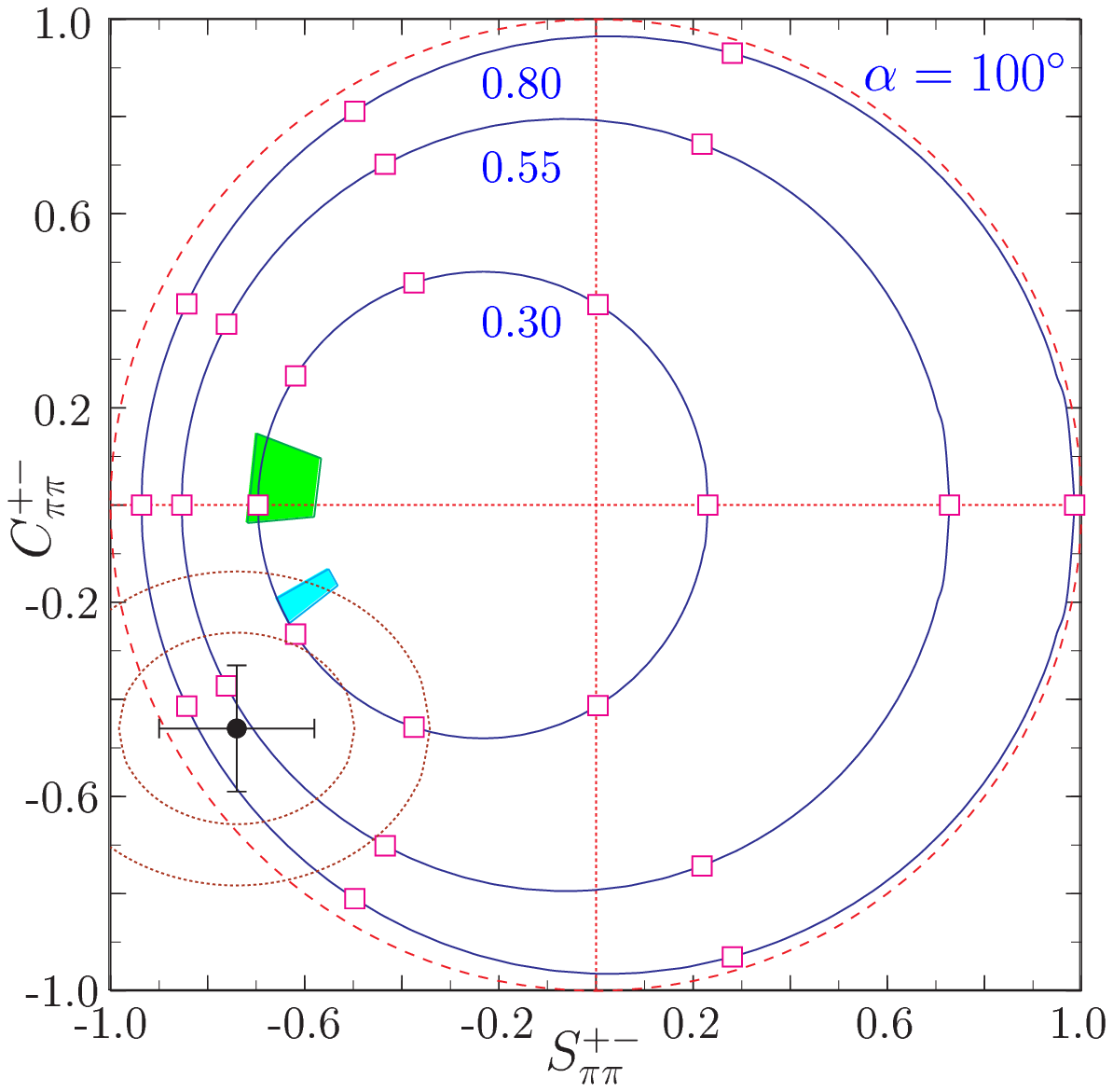}
\quad
\psfig{width=0.40\textwidth,file=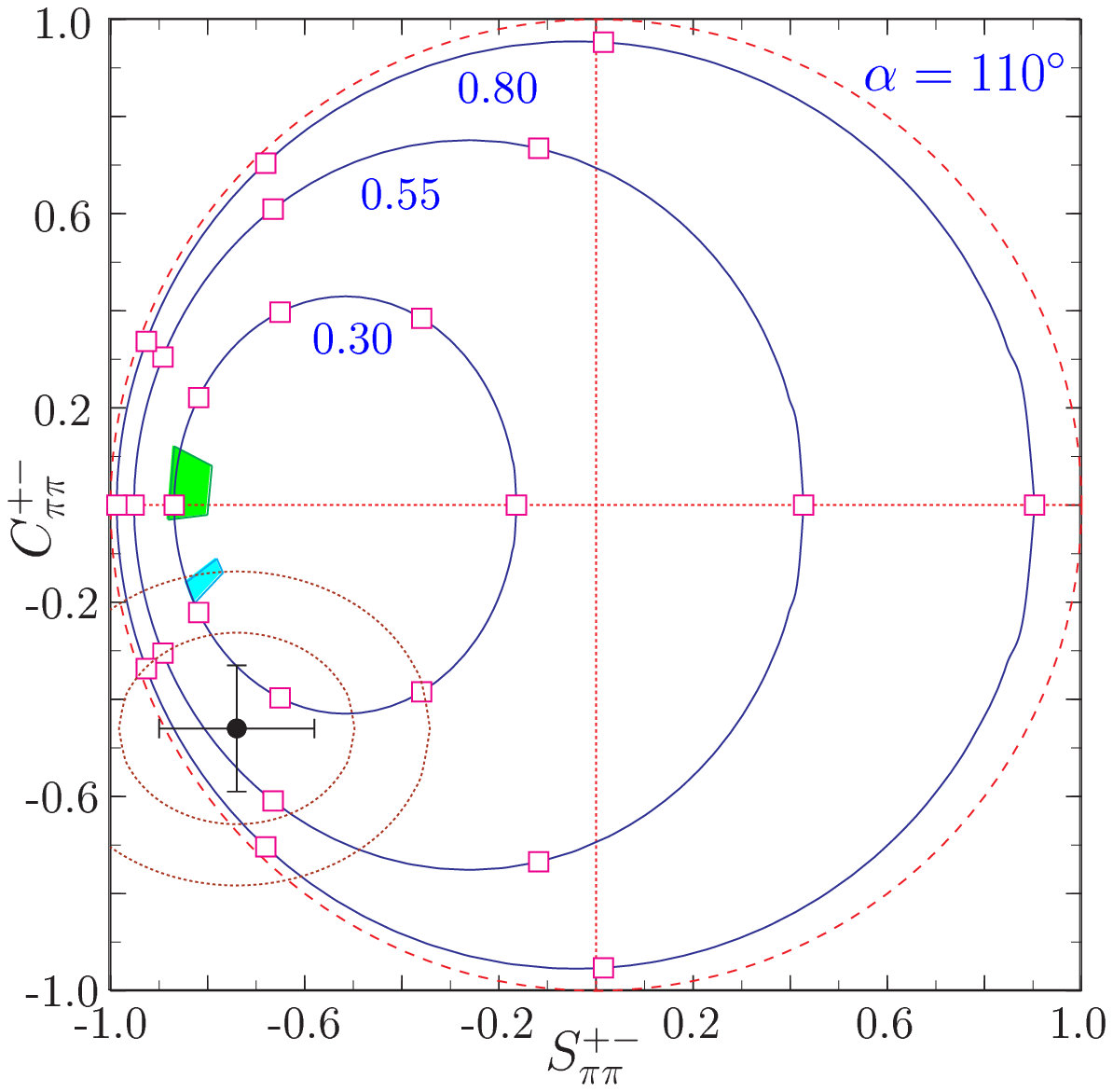}
}
\bigskip
\centerline{
\psfig{width=0.40\textwidth,file=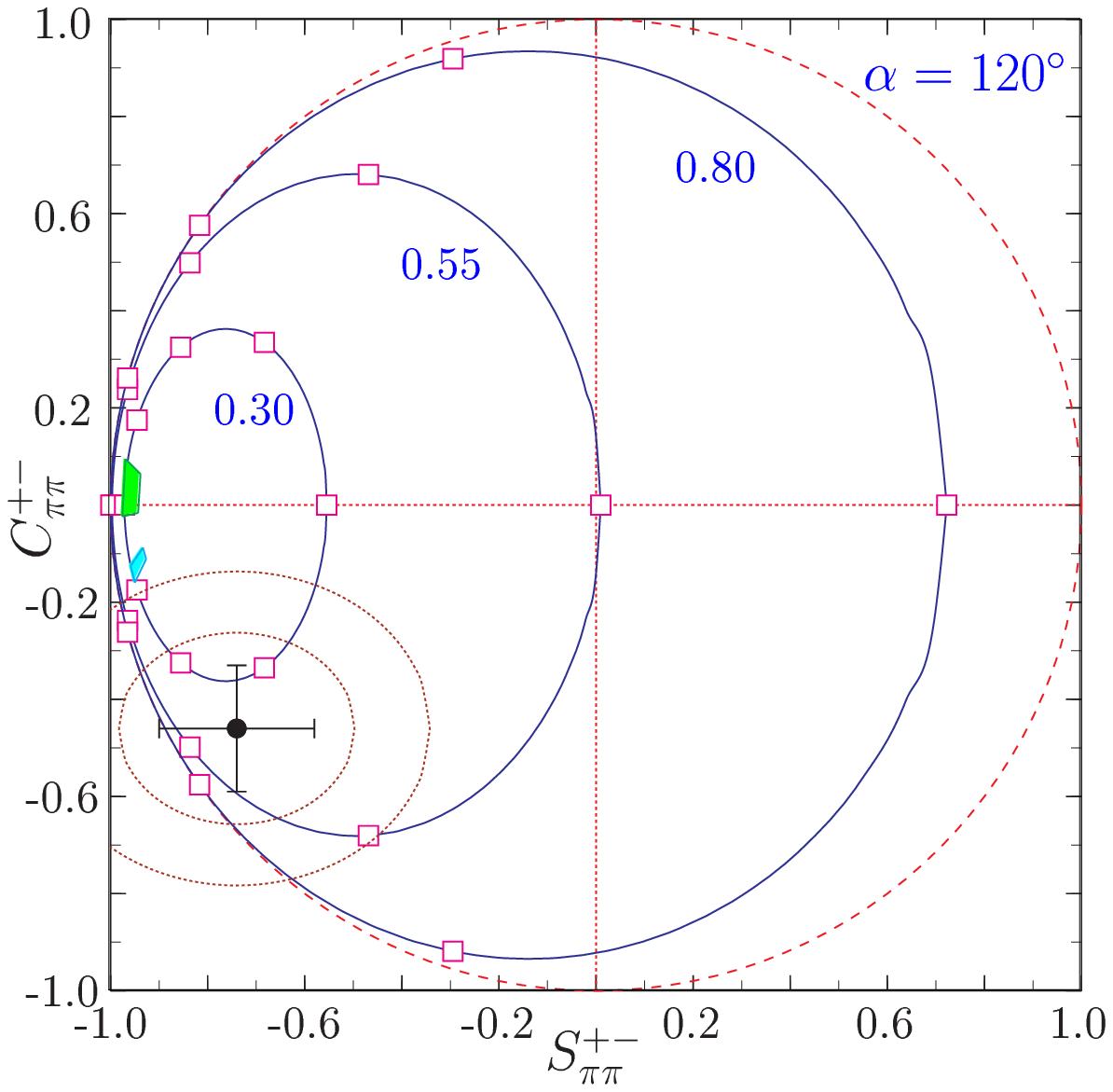}
\quad
\psfig{width=0.40\textwidth,file=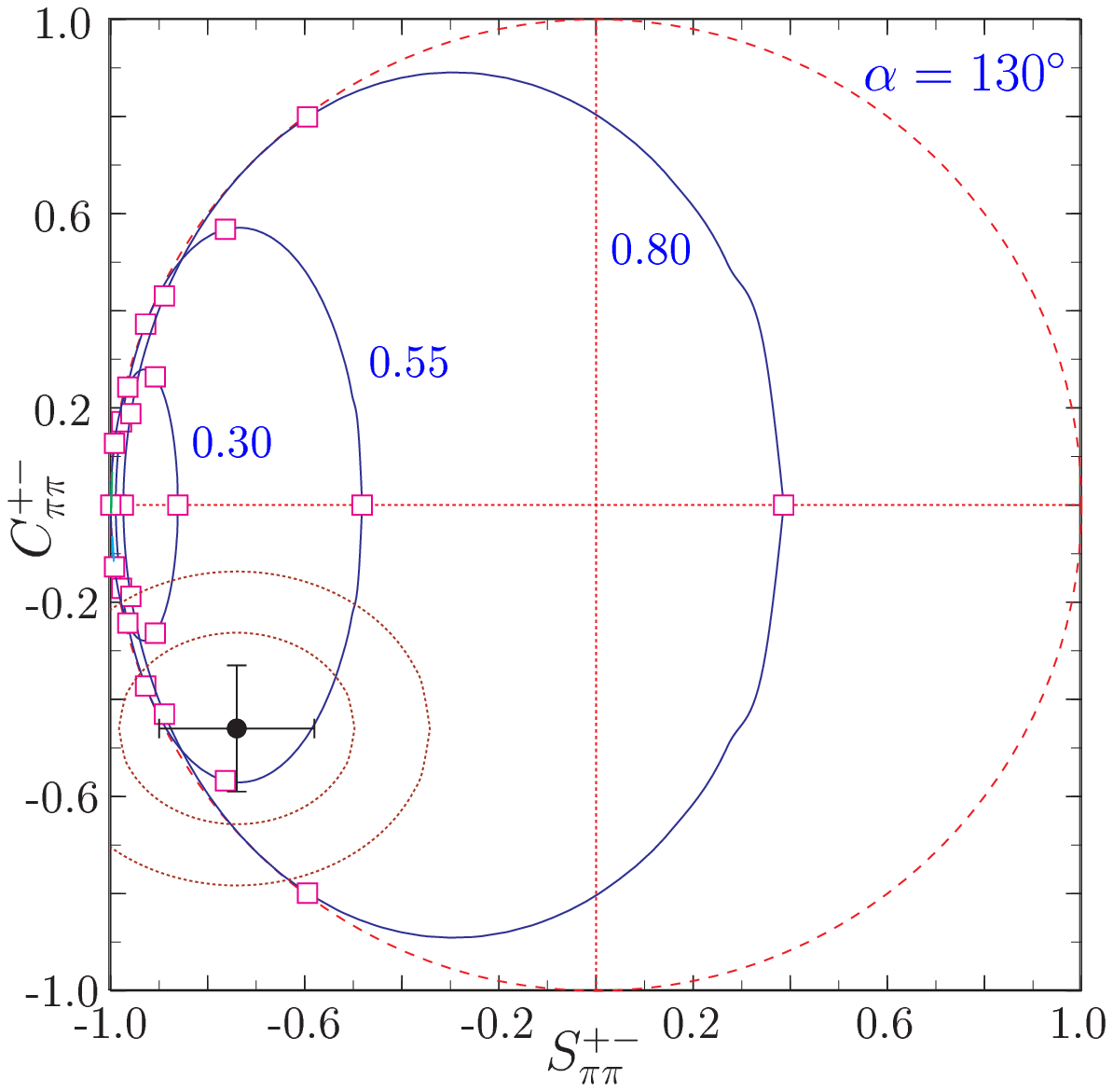}
}
\caption{Implications of the time-dependent CP asymmetry parameters
         $C_{\pi \pi}^{+-} = -0.46 \pm 0.13$ and
         $S_{\pi \pi}^{+-} = -0.74 \pm 0.16$ from the BELLE and BABAR
         measurements for the CP violating phase~$\alpha$ (or~$\phi_2$).
         In this analysis, the strong phase~$\delta_c$ is varied over
         the full range $-\pi \leq \delta_c \leq \pi$ and curves 
         are drawn for three values $|P_c/T_c| = 0.30$, $0.55$, and~$0.80$. 
         The predictions of the QCD factorization (upper box) and pQCD
         (lower box) approaches are also shown for fixed values of~$\alpha$
         noted on the six frames. The curves around the data point 
         represent the 68.3\% and 95.5\%~C.L. contours.}
\label{fig:B-pipi}
\end{figure}

We start by showing that the current data on~$C_{\pi\pi}^{+-}$
and~$S_{\pi\pi}^{+-}$ in the $B^0_d \to \pi^+ \pi^-$ decays provides 
a discrimination among various dynamical approaches, for which 
the QCD factorization~\cite{Beneke:1999br} and perturbative 
QCD~\cite{Keum:2000ph} approaches will be taken as the two leading
contenders.  The results of this analysis are 
presented in Fig.~\ref{fig:B-pipi} for six values of~$\alpha$ 
in the range $80^\circ \leq \alpha \leq 130^\circ$ in intervals 
of~$10^\circ$. To take into account the dispersion in the values  
of~$|P_c /T_c|$, we take three values of this ratio, namely~$0.30$, 
$0.55$, and~$0.80$. The first of these values represents the current 
expectations of this quantity, whereas the last is taken with the 
hindsight of the best fit of the data that we have performed in 
a model-independent way, as described later.
The points indicated on these contours represent 
the values of the strong phase difference~$\delta_c$ which is varied 
in the interval $-\pi \leq \delta_c \leq \pi$. We do not show the 
plot for $\alpha = 70^\circ$, which is the 95\%~C.L. lower value
of~$\alpha$ from the unitarity fits, as already the case 
$\alpha = 80^\circ$ requires rather large value of~$|P_c/T_c|$. 
In each figure, the outer circle corresponds to the constraint 
$(S_{\pi \pi}^{+-})^2 + (C_{\pi \pi}^{+-})^2 = 1$. The current 
average~(\ref{eq:Spipi-Cpipi-av-new}) of the BABAR and BELLE data 
satisfies this constraint as shown by the data point with (unscaled) 
errors. The two ellipses surrounding the experimental measurement
represent the~68.3\% and~95.5\%~C.L. contours.
This figure demonstrates that, as $C_{\pi \pi}^{+-}$ is 
negative and large, current data favors a rather large strong 
phase, typically $-60^\circ \leq \delta_c \leq -30^\circ $. 
The two shaded regions shown in this figure correspond to the
predictions of the QCD-factorization approach (the upper shaded 
area) and the perturbative QCD framework (the lower shaded area). 
As can be seen, the predictions of the QCD-factorization approach 
lie outside of the~$3\sigma$ experimental 
measurements for all values of~$\alpha$ shown in this figure. 
For the  perturbative QCD framework~\cite{Keum:2000ph}, one finds 
agreement with the measurements, but only at about~$2 \sigma$ level. 
Restricting $\alpha$ in the region $90^\circ \leq \alpha \leq 110^\circ$,
good fits of the data are obtained for typically $|P_c /T_c|\geq 0.5$ and
$\delta_c \le - 30^\circ$. We shall quantify the fits more precisely later.

%
%

\begin{figure}[tb]
\centerline{
\psfig{width=0.95\textwidth,file=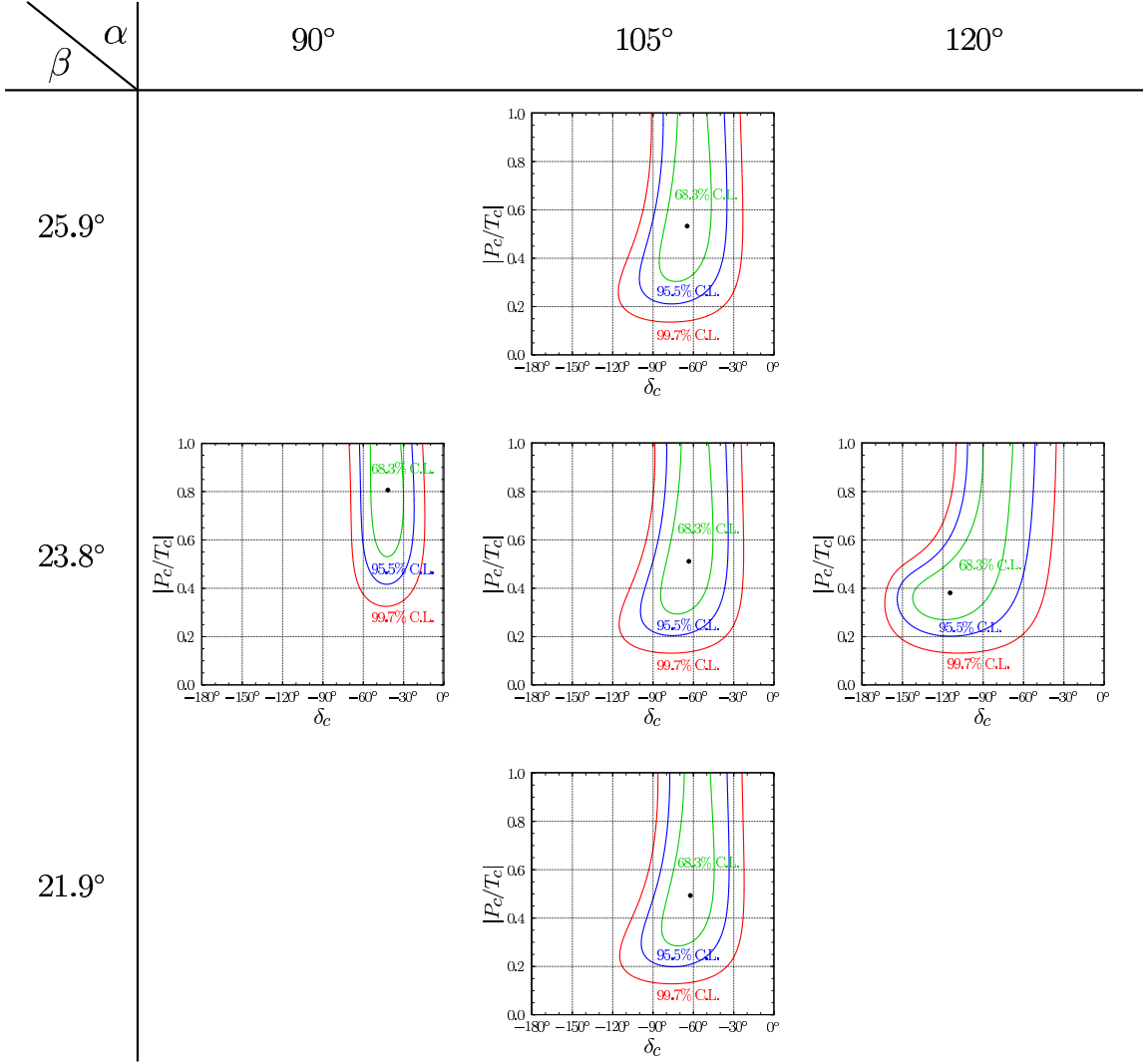}
}
\caption{The correlation $|P_c/T_c| - \delta_c$ corresponding 
         to the 68.3\%, 95.5\%, and 99.7\%~C.L. ranges 
         of~$C_{\pi \pi}^{+-}$ and~$S_{\pi \pi}^{+-}$ for three 
         values of~$\alpha$ and~$\beta$. Note that the dependence 
         on~$\beta$ is rather weak and hence not shown for the 
         other two values of~$\alpha$.}
\label{fig:PT-delta}
\end{figure}

\begin{figure}[tb]
\centerline{
\psfig{width=0.95\textwidth,file=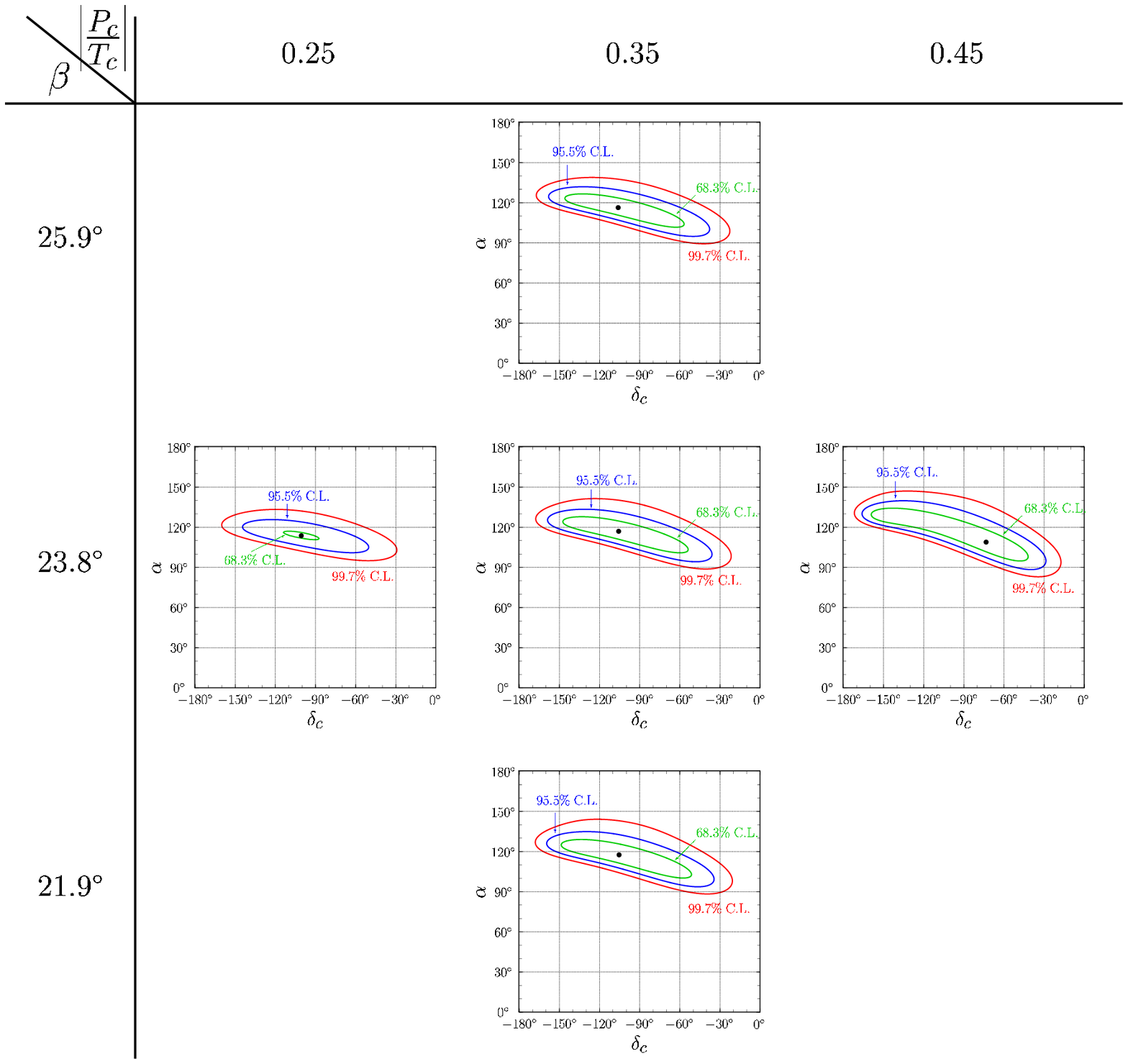}
}
\caption{The correlation $\alpha - \delta_c$ corresponding 
         to the 68.3\%, 95.5\%, and 99.7\%~C.L. ranges 
         of~$C_{\pi \pi}^{+-}$ and~$S_{\pi \pi}^{+-}$ for 
         $\vert P_c/T_c \vert= 0.25$, $0.35$, and~$0.45$.}
\label{fig:alpha-delta-1}
\end{figure}

\begin{figure}[tb]
\centerline{
\psfig{width=0.95\textwidth,file=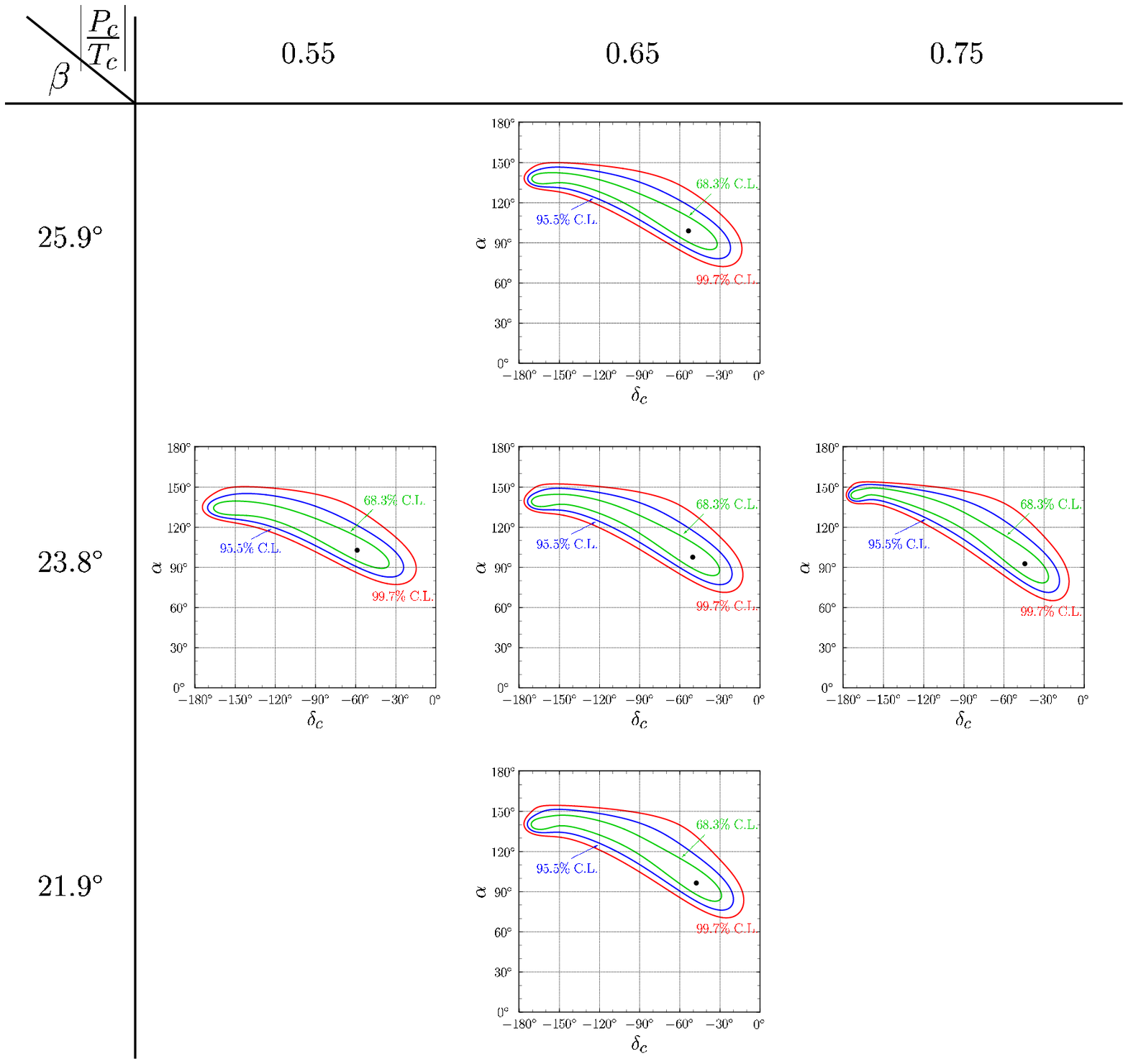}
}
\caption{The correlation $\alpha - \delta_c$ corresponding 
         to the 68.3\%, 95.5\%, and 99.7\%~C.L. ranges 
         of~$C_{\pi \pi}^{+-}$ and~$S_{\pi \pi}^{+-}$ for  
         $\vert P_c/T_c \vert = 0.55$, $0.65$, and~$0.75$.}
\label{fig:alpha-delta-2}
\end{figure}

We now turn to a model-independent analysis of the~$C_{\pi\pi}^{+-}$
and~$S_{\pi\pi}^{+-}$ data. As the current world averages of these 
quantities are negative and rather large~(\ref{eq:Spipi-Cpipi-av-new}), 
positive values of the strong phase difference~$\delta_c$ are excluded 
at a high confidence level ($>99.7\%$~C.L.) which is demonstrated 
in Figs.~\ref{fig:PT-delta}, \ref{fig:alpha-delta-1} 
and~\ref{fig:alpha-delta-2}. This observation is the reason why the 
full range $-\pi \leq \delta_c \leq +\pi$ has been  restricted to 
the negative values of~$\delta_c$, $-\pi \leq \delta_c \leq 0$,
in these figures. In Fig.~\ref{fig:PT-delta}, three representative 
values ($90^\circ$, $105^\circ$ and~$120^\circ$) of the angle~$\alpha$ 
from the UT-favored interval~(\ref{eq:phi13-ckm-fitter}) and three 
values ($25.9^\circ$, $23.8^\circ$ and~$21.9^\circ$) of the angle~$\beta$, 
which cover the present measurement of this quantity within~$\pm 1 \sigma$ 
range~(\ref{eq:sin2beta-exp}), are shown and the resulting 
$\chi^2$-contours in the variables~$|P_c/T_c|$ and~$\delta_c$ are plotted.  
Note that the dependence on the precise value of~$\beta$ in the current 
experimental range of this angle is rather weak. Hence, we show the
$\beta$-dependence of the correlation for only one value of~$\alpha$, 
namely $\alpha = 105^\circ$.  The most important message from this analysis 
is that the current data favours negative and rather large values of 
the strong phase~$\delta_c$, which are correlated with the values 
of~$\alpha$. Restricting to the 68.3\%~C.L. contours for the sake of 
definiteness, the minimum allowed values of~$-\delta_c$ are:~$30^\circ$, 
$45^\circ$, and~$70^\circ$ for $\alpha = 90^\circ$, $105^\circ$, 
and~$120^\circ$, respectively. What concerns the allowed values
of $|P_c/T_c|$, we note that except for a relatively small allowed 
region near $80^\circ \leq \alpha \leq 90^\circ$, they 
overlap with the theoretical estimates of the same specified above
at 95.5\%~C.L. However, the best-fit values of~$|P_c/T_c|$ are on the 
higher side as shown by the dots in these figures. It should be noted 
that the current data results in the lower bound on the penguin-to-tree 
ratio $|P_c/T_c| \geq 0.18$ at 95.5\%~C.L but extends to much larger 
values of~$|P_c/T_c|$, which we have suppressed in these figures for 
the sake of clarity but will show 
in the next section where 
we discuss the fits of the unitarity triangle.

The correlations between~$\alpha$ and~$\delta_c$ for three fixed 
values~$|P_c/T_c| = 0.25$, $0.35$, and~$0.45$ in the theoretically 
motivated interval are shown in Fig.~\ref{fig:alpha-delta-1}. This 
figure updates the results by the BELLE collaboration~\cite{Abe:2003ja} 
and shows that a satisfactory description of the current data for 
these values of~$|P_c/T_c|$ and with~$\alpha$ lying within the 
indirect UT-based range is possible only with large values of the 
strong phase~$-\delta_c$. Again using the 68.3\%~C.L. contours, 
it is seen that the minimum allowed value for $|P_c/T_c| = 0.35$ 
is $-\delta_c \simeq 55^\circ$, and it decreases to~$45^\circ$ 
for $|P_c/T_c| = 0.45$. With $0.20 < |P_c/T_c| < 0.45$, the 
angles~$\alpha$ and~$\delta_c$ lie in the intervals: 
$90^\circ \leq \alpha \leq 130^\circ$ and 
$-160^\circ \leq \delta_c \leq -30^\circ$, at the 95.5\%~C.L.
As higher values of~$|P_c/T_c|$ are experimentally allowed, 
the correlations between~$\alpha$ and~$\delta_c$ for larger values 
of~$|P_c/T_c| = 0.55$, $0.65$, and~$0.75$ are shown in 
Fig.~\ref{fig:alpha-delta-2}. We note that with these values,
the allowed ranges for the angles~$\alpha$ and~$\delta_c$  
become wider with increasing $|P_c/T_c|$.  

\begin{figure}[htb]
\centerline{
\psfig{width=0.95\textwidth,file=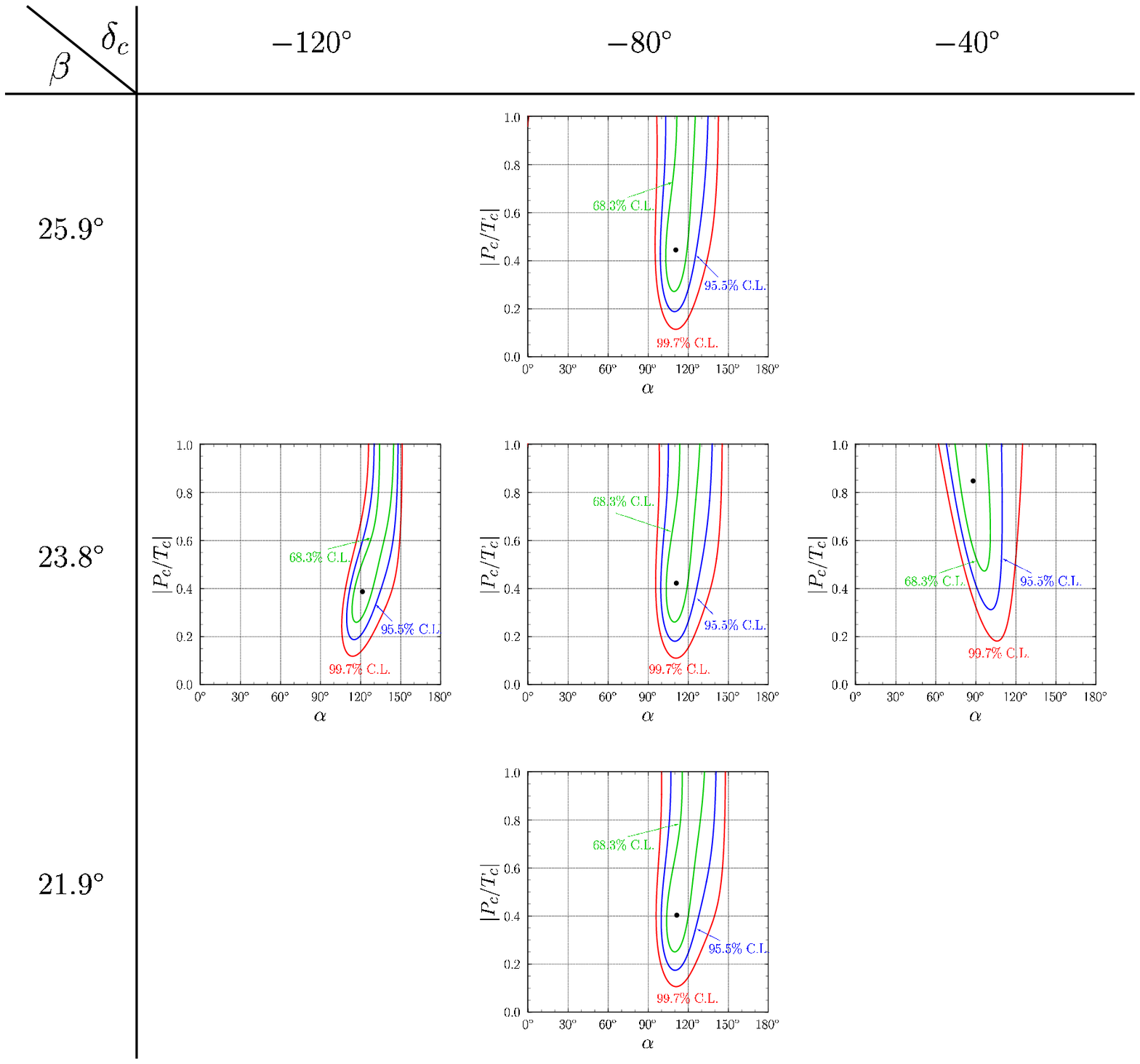}
}
\caption{The  correlation ~$|P_c/T_c| - \alpha$  corresponding    
         to the 68.3\%, 95.5\%, and 99.7\%~C.L. ranges 
         of~$C_{\pi \pi}^{+-}$ and~$S_{\pi \pi}^{+-}$ for
         $\delta_c = -120^\circ$, $-80^\circ$, and~$-40^\circ$.}
\label{fig:PT-alpha}
\end{figure}

The correlations between the angle~$\alpha$ and~$|P_c/T_c|$,
 for three representative values of the 
strong-phase difference $\delta_c = -40^\circ$, $-80^\circ$, 
and~$-120^\circ$ and the angle~$\beta$ within its experimental 
range, are presented in Fig.~\ref{fig:PT-alpha}. This figure 
demonstrates again that smaller values of~$|\delta_c|$ require 
larger values of~$|P_c/T_c|$. The restrictions on~$|P_c/T_c|$ 
and~$\alpha$ discussed above are also seen in this figure.
 
In summary, we see that current data allows a wide range of the
quantities $|P_c/T_c|$ and $\delta_c$, and without restricting 
them the impact of the~$C_{\pi\pi}^{+-}$ and~$S_{\pi\pi}^{+-}$ 
measurements on the CKM parameters, in particular the angles~$\alpha$ 
or~$\gamma$, is rather small. The dynamical approaches discussed
above are not a great help as they are not good fits of the data
within the SM.

%
%

%
%
%

%
%
%

%
%

\subsection{Constraints on $\cos (2\theta)$ from bounds based 
            on the isospin symmetry}

\begin{figure}[tbp!]
\centerline{
\psfig{width=0.40\textwidth,file=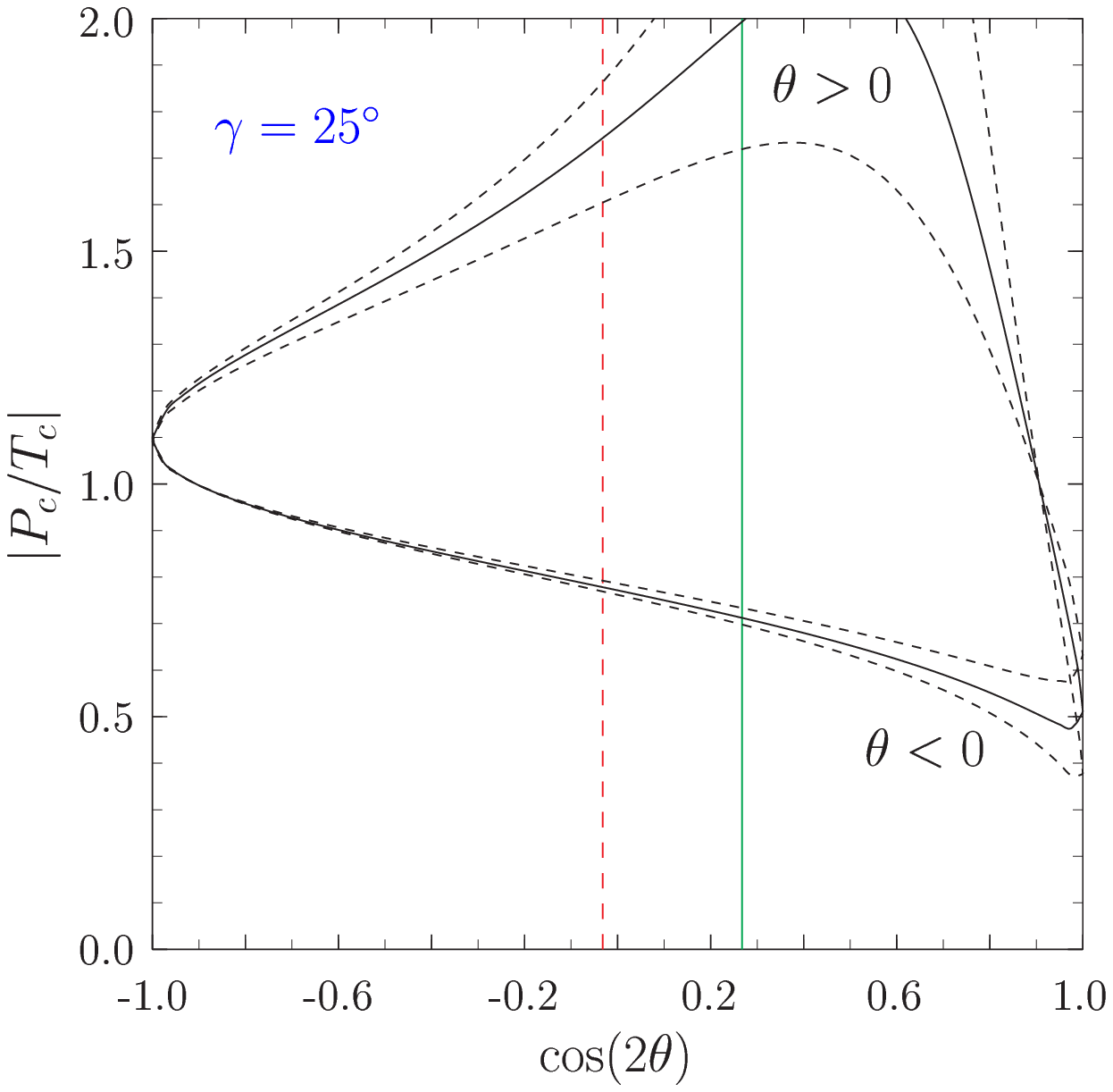}
\quad
\psfig{width=0.40\textwidth,file=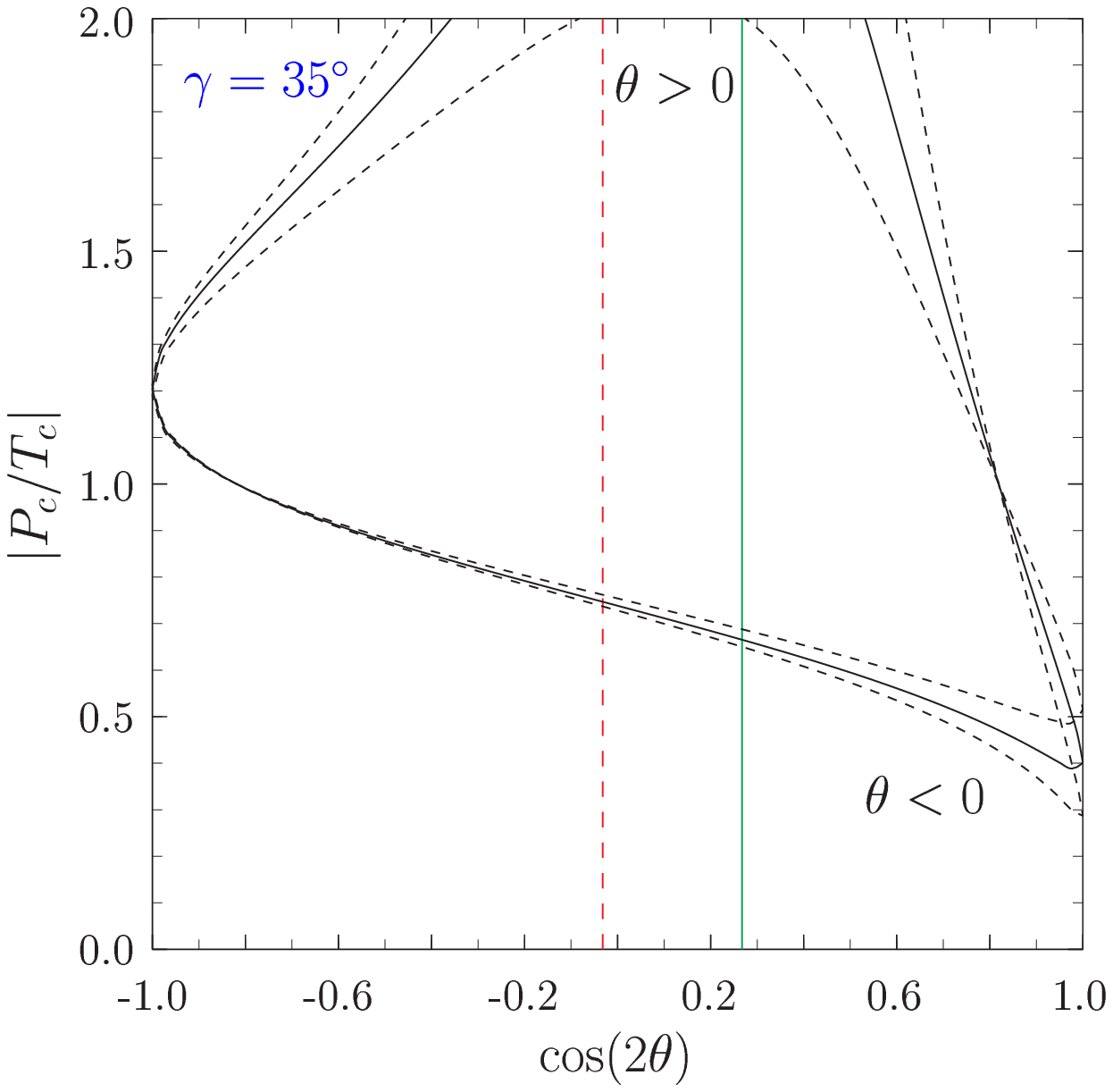}
}
\bigskip
\centerline{
\psfig{width=0.40\textwidth,file=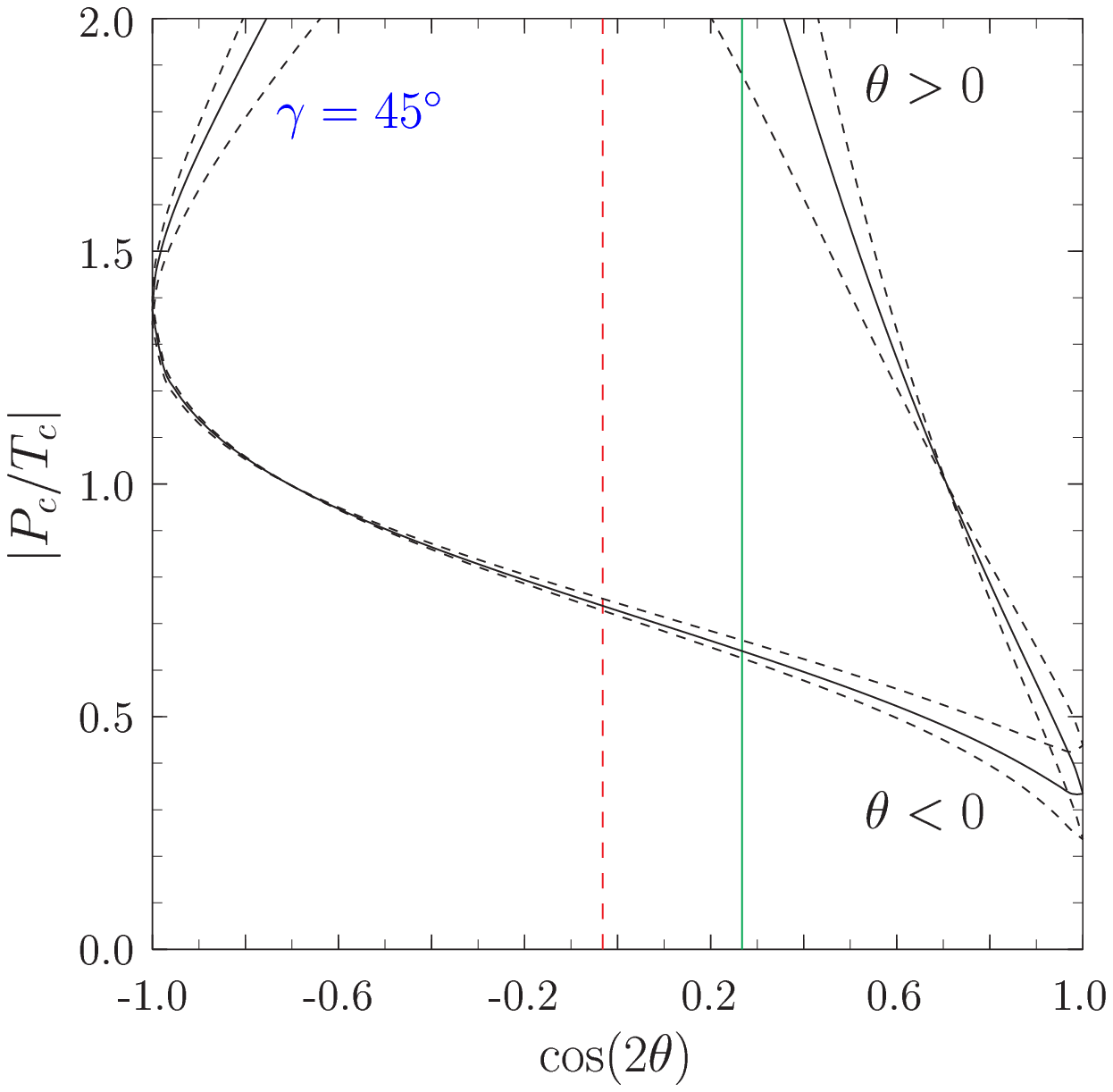}
\quad
\psfig{width=0.40\textwidth,file=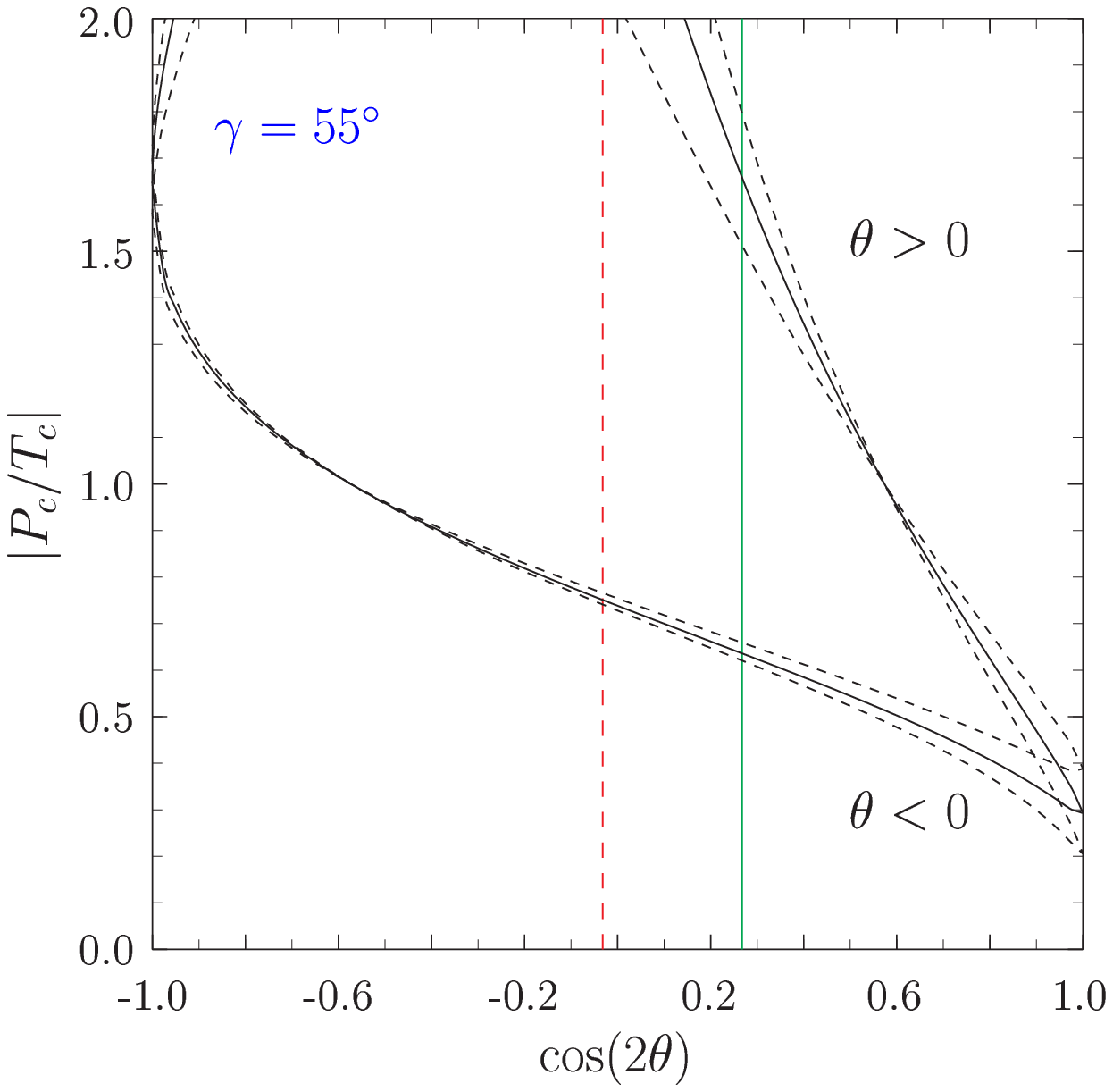}
}
\bigskip
\centerline{
\psfig{width=0.40\textwidth,file=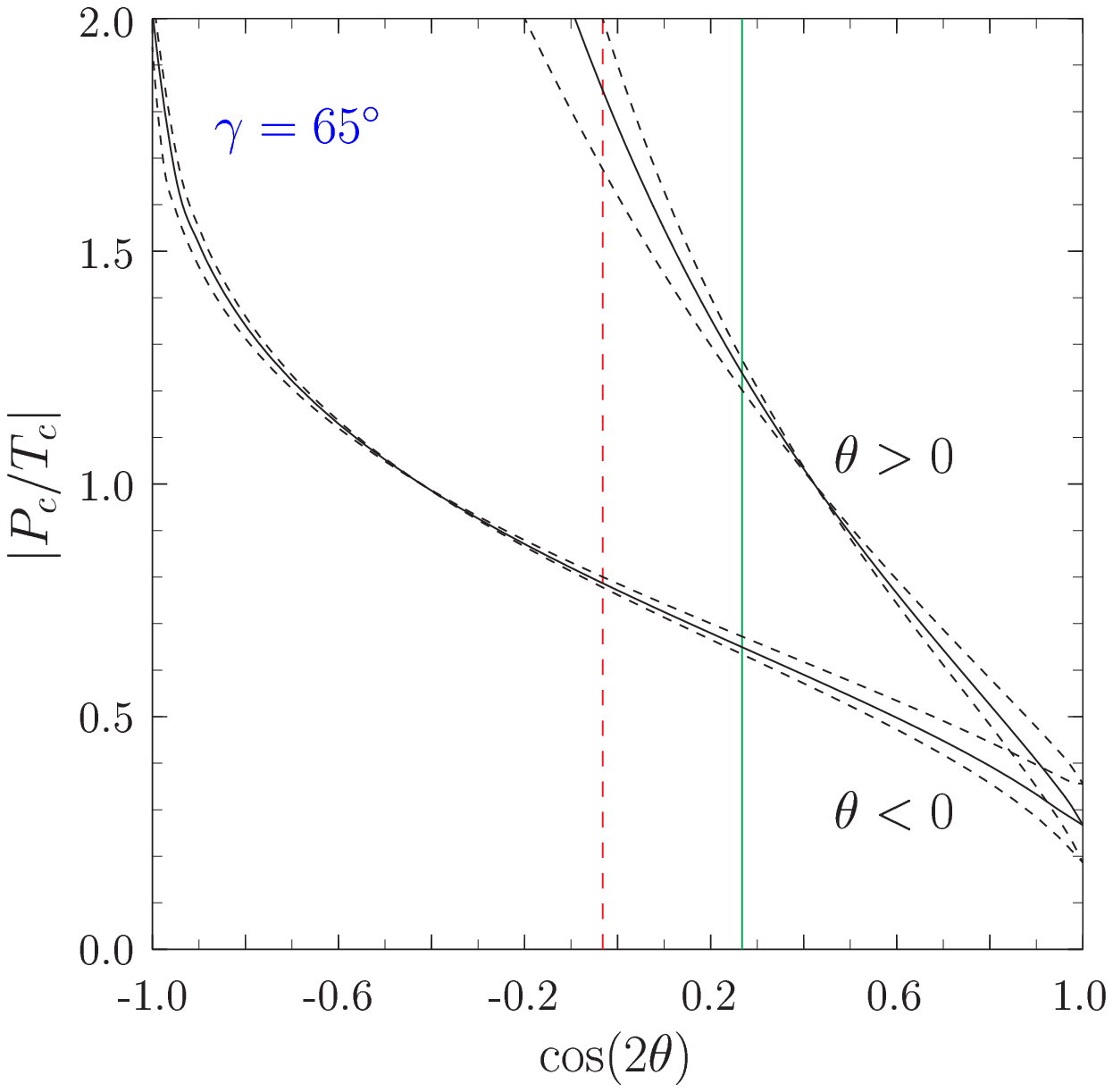}
\quad
\psfig{width=0.40\textwidth,file=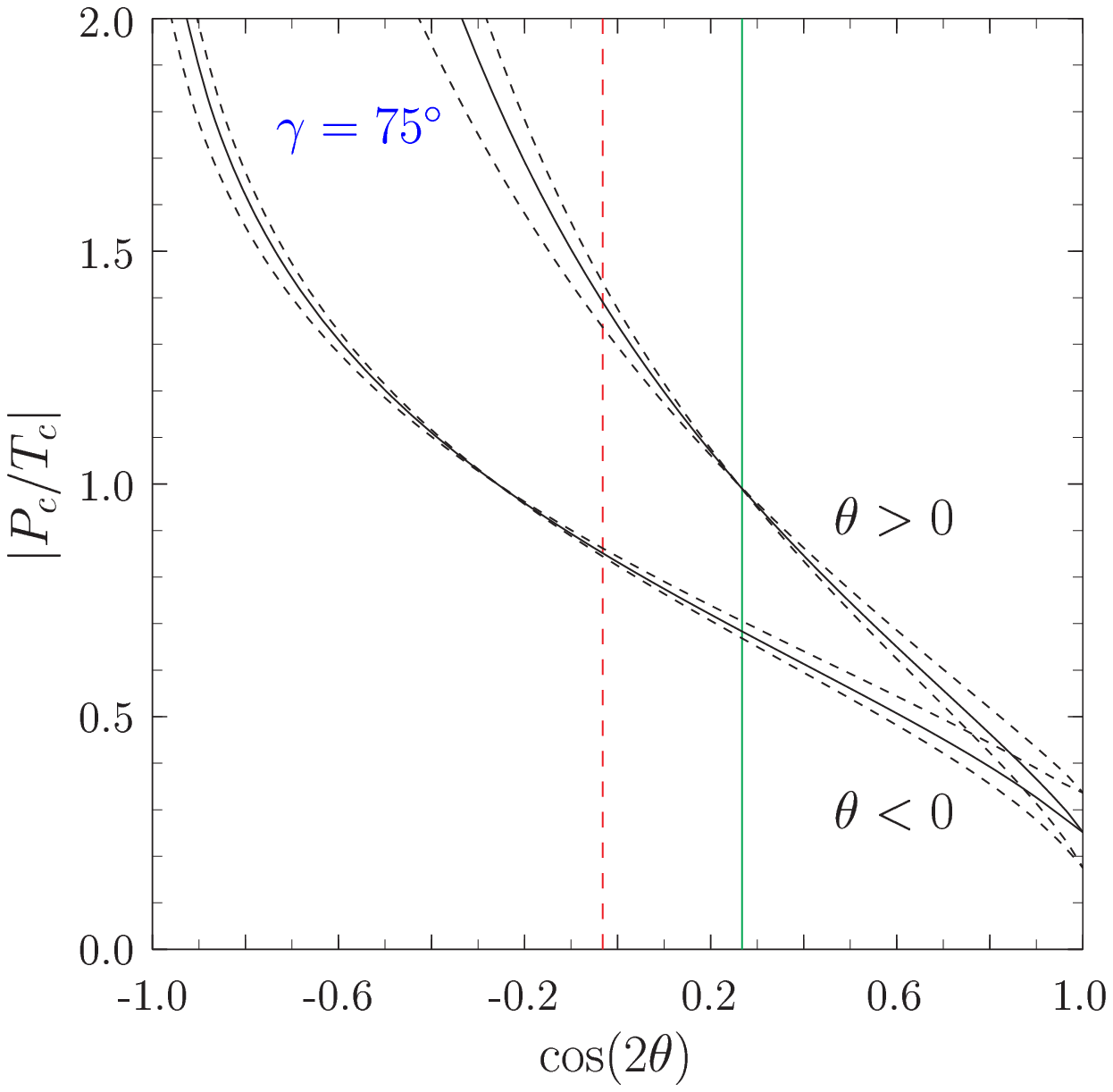}
}
\caption{The correlation $|P_c/T_c| - \cos (2\theta)$ based on  
         the current measurements of~$C_{\pi \pi}^{+-}$ for fixed  
         values of the angle~$\gamma$ indicated on the frames. 
         The red dashed and green solid vertical lines represent  
         the Gronau-London-Sinha-Sinha (GLSS) lower bounds   
         $\cos (2\theta)> -0.03$ and $\cos (2\theta)> 0.27$, 
         respectively, as discussed in the text.
         The upper and lower sets     
         of curves correspond to the branches with $\theta > 0$  
         and $\theta < 0$, respectively.}
\label{fig:PT-Cos2T}
\end{figure}

As discussed in the previous section, the penguin contribution 
in the $B^0_d \to \pi^+ \pi^-$ decay can be parameterized by the 
angle~$2\theta$. Having at hand the experimental range 
of~$C_{\pi \pi}^{+-}$, it is of interest to work out numerically 
the dependence between the ratio~$|P_c/T_c|$ and~$\cos(2\theta)$, 
given by Eq.~(\ref{eq:PcTc2}). The results of this 
analysis are presented in Fig.~\ref{fig:PT-Cos2T} for six values 
of~$\gamma$ in the range $25^\circ \leq \alpha \leq 75^\circ$ 
in intervals of~$10^\circ$. The solid lines in all the frames 
correspond to the central experimental value 
of~$C_{\pi \pi}^{+-}$~(\ref{eq:Spipi-Cpipi-av-new}) 
while the dashed lines correspond to the $\pm 1 \sigma$ values of 
this quantity. Due to the functional dependence~(\ref{eq:PcTc2})
of~$r_c^2$ on $\cos (2 \theta - 2 \gamma)$, there exists a
sign ambiguity and there are two solutions depending on~$\theta > 0$ 
and~$\theta < 0$. We show both of these solutions and each frame 
in this figure contains two sets of curves where the upper 
and the lower ones correspond to $\theta > 0$ and $\theta < 0$, 
respectively. The isospin symmetry and the existing data on the 
$B \to \pi \pi$ decays allow to put restrictions on $\cos(2\theta)$. 
The GLSS lower bound~(\ref{eq:GLSS-bound}) on $\cos(2\theta)$ is 
based on the $B \to \pi \pi$ branching ratios and $y_{\pi\pi}^{+-}$. 
The recent experimental data on the branching ratios and the~$B^+$- 
and $B^0$-meson lifetime ratio~\cite{HFAG:2004}:  
\begin{eqnarray} 
{\cal B} (B^0_d \to \pi^+ \pi^-) & = & (4.55 \pm 0.44) \times 10^{-6} , 
\nonumber \\ 
{\cal B} (B^+ \to \pi^+ \pi^0) & = & (5.27 \pm 0.79) \times 10^{-6} , 
\nonumber \\ 
{\cal B} (B^0_d \to \pi^0 \pi^0) & = & (1.90 \pm 0.47) \times 10^{-6} , 
\nonumber \\  
\tau_{B^+}/\tau_{B^0} & = & 1.086 \pm 0.017 , 
\nonumber 
\end{eqnarray}
have been used in getting the conservative numerical bound:
$\cos(2\theta) > - 0.03$.
This bound is shown as vertical dashed lines in all the frames 
in Fig.~\ref{fig:PT-Cos2T}. It should be noted that if the 
central values of the data are used instead, the resulting 
GLSS bound is:
\begin{equation} 
\cos(2\theta) \Big |_{\rm GLSS} > 0.27~,
\label{eq:GLSS-restriction}
\end{equation}
which is shown as the solid vertical lines in  
Fig.~\ref{fig:PT-Cos2T}.  The shift is mainly due to the current 
uncertainties in the branching ratios for $B^0_d \to \pi^0 \pi^0$ 
and $B^+ \to \pi^+ \pi^0$. Our analysis shows that putting a lower 
bound on $\cos (2\theta)$,
$|P_c/T_c|$ gets significantly constrained. It is seen that 
the branch with $\theta < 0$ results in smaller values for~$|P_c/T_c|$,
which are concentrated in a relatively narrow interval. However, 
as~$\gamma$ increases, this interval becomes wider. {\it A~priori}, 
it is difficult to argue which of the two solutions $\theta >0$ and 
$\theta < 0$ should be entertained. Hence, in the implementation of 
the isospin-based bound on $\cos (2\theta)$ in the unitarity fits, 
we shall allow the sign of~$\theta$ to take either value.

Based on the central values of the experimental data 
specified above and $C_{\pi\pi}^{+-}$~(\ref{eq:Spipi-Cpipi-av-new}), 
the minimal value of the direct CP asymmetry in the 
$B^0_d \to \pi^0 \pi^0$ decay~(\ref{eq:Cpipi00-GLSS-bound}) 
can be estimated as: 
\begin{equation}
C_{\pi\pi}^{00} \ge 0.47. 
\label{eq:GLSS-restriction-ACPdir00}
\end{equation}
It should be noted that~$C_{\pi\pi}^{00}$ differs in sign 
from~$C_{\pi\pi}^{+-}$. 
(See, also the recent analysis by Buras et al.~\cite{Buras:2004ub}.) 
%

%
%

%
\begin{figure}[tb]
\centerline{
\psfig{width=0.50\textwidth,file=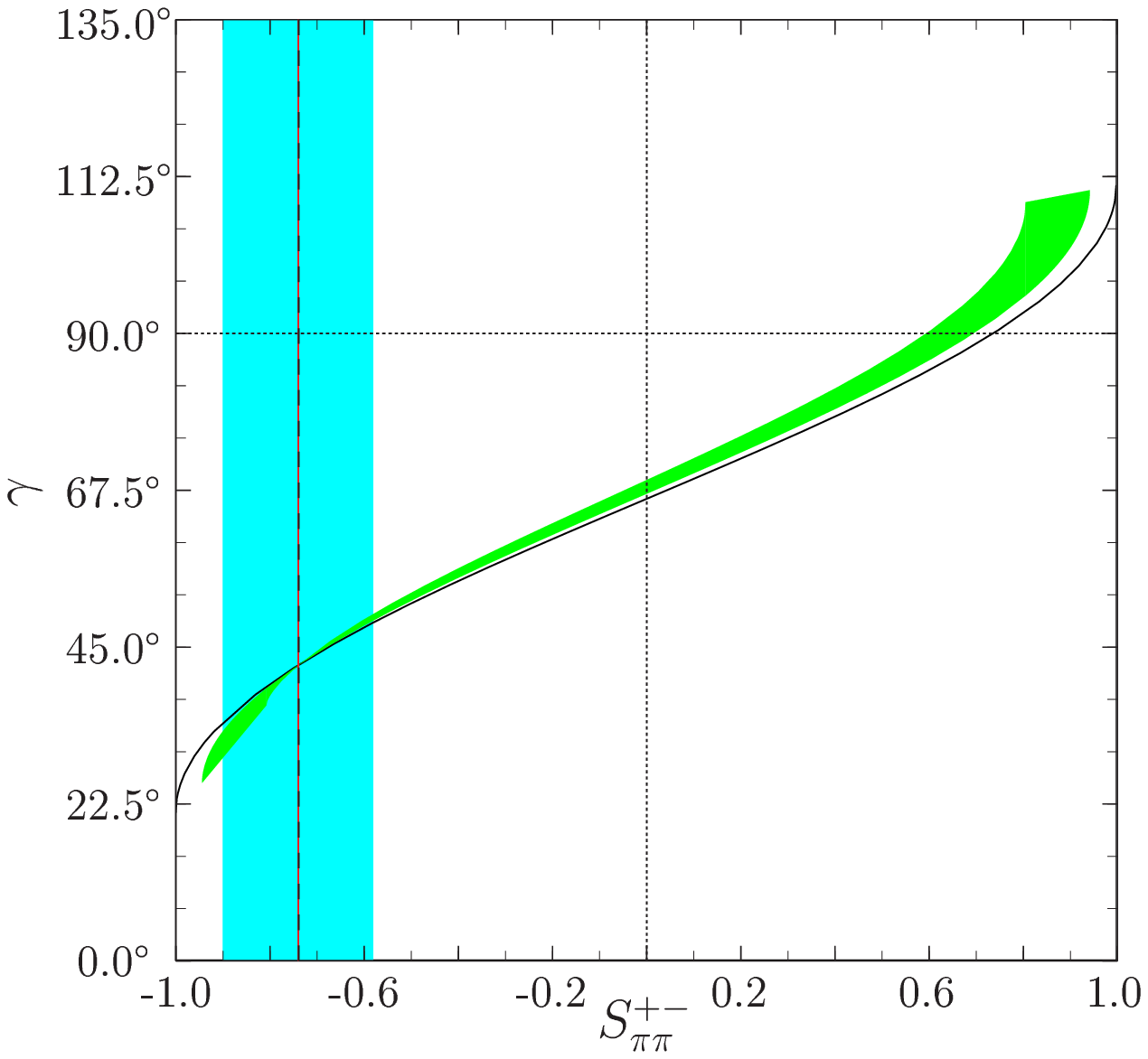}
}
\caption{
The SM limits on the angle~$\gamma$ in dependence on
$S_{\pi\pi}^{+-}$ at $C_{\pi \pi}^{+-} = 0$ (the Buchalla-Safir
limit) and at $C_{\pi \pi}^{+-} = -0.46 \pm 0.13$
(the Botella-Silva limit) shown as the solid line and
the shaded area, respectively. Note that both are the lower
limit for $S_{\pi\pi}^{+-} > -\sin(2\beta)$ and the upper
one for $S_{\pi\pi}^{+-} < -\sin(2\beta)$. The vertical band
corresponds to the experimentally measured value 
of~$S_{\pi\pi}^{+-}$ with the recent central value which 
practically coincides with $S_{\pi\pi}^{+-} = - \sin(2\beta)$.
}
\label{fig:gam-Spipi}
\end{figure}
%

The SM-based bounds  on the angle~$\gamma$ as a function of the 
CP asymmetry~$S_{\pi\pi}^{+-}$ with $C_{\pi \pi}^{+-} = 0$ (the
Buchalla-Safir bound~\cite{Buchalla:2003jr}) and with 
$C_{\pi \pi}^{+-} = -0.46 \pm 0.13$ (the Botella-Silva 
bound~\cite{Botella:2003xp}) are shown in Fig.~\ref{fig:gam-Spipi} 
as the solid line and the shaded area, respectively. The vertical 
band corresponds to the current experimentally measured value, 
and the central value practically coincides with 
$S_{\pi\pi}^{+-} = - \sin (2\beta)$. Thus, these limits do not   
provide any restrictions on $\gamma$ at present, but if with 
improved data a sizable shift of the~$S_{\pi\pi}^{+-}$ central 
value from its current value takes place, then these bounds may 
lead to useful constraints.

\section{Analysis of the CKM Unitarity Triangle Including
         $C_{\pi\pi}^{+-}$ and $S_{\pi\pi}^{+-}$ Measurements}
\label{sec:ut}

In this section we investigate the impact of the~$C_{\pi\pi}^{+-}$ 
and~$S_{\pi\pi}^{+-}$ measurements on the unitarity triangle fits. 
We adopt a bayesian analysis method to fit the data. Systematic and 
statistical errors are combined in quadrature. We add a contribution 
to the chi-square for each of the inputs presented in
Table~\ref{inputparms}. Other input quantities are taken from their
central values given in the PDG review~\cite{Hagiwara:fs}.
The lower bound on~$\Delta M_{B_s}$ is
implemented using the modified-$\chi^2$ method (as described in the
CERN CKM Workshop proceedings~\cite{Battaglia:2003in}), which makes 
use of the amplitude technique~\cite{Moser:1996xf}. 
The $B_s \leftrightarrow \bar B_s$ oscillation probabilities are 
modified to have the dependence $P(B_s \to \bar B_s) \propto 
[1 + {\cal A} \cos (\Delta M_{B_s} \, t)]$ and $P(B_s \to B_s) 
\propto [1 - {\cal A} \cos (\Delta M_{B_s} \, t)]$. 
The contribution to the $\chi^2$-function is then
\begin{equation}
\chi^2 (\Delta M_{B_s}) = 2 \left[
{\rm Erfc}^{-1} \left( {1\over 2} \, {\rm Erfc} \; 
{1 - {\cal A} \over \sqrt{2} \, \sigma_{\cal A}}
\right) \right]^2 \,, 
\label{eq:chi2-DMBs} 
\end{equation}
where~${\cal A}$ and~$\sigma_{\cal A}$ are the world average amplitude 
and error, respectively. The measurements of~$C_{\pi\pi}^{+-}$ 
and~$S_{\pi\pi}^{+-}$ contribute to the $\chi^2$-function according to 
Eq.~(\ref{eq:chi2}). The resulting $\chi^2$-function is then minimized 
over the following parameters: $\bar \rho$, $\bar \eta$, $A$, $\hat B_K$, 
$\eta_1$, $\eta_2$, $\eta_3$, $m_c (m_c)$, $m_t (m_t)$, $\eta_B$, 
$f_{B_d} \sqrt{B_{B_d}}$, $\xi$, $|P_c/T_c|$, and~$\delta_c$. Further
details can be found in Ref.~\cite{Ali:2003te}. 

\begin{figure}[htb]
\centerline{
\psfig{width=1.00\textwidth,file=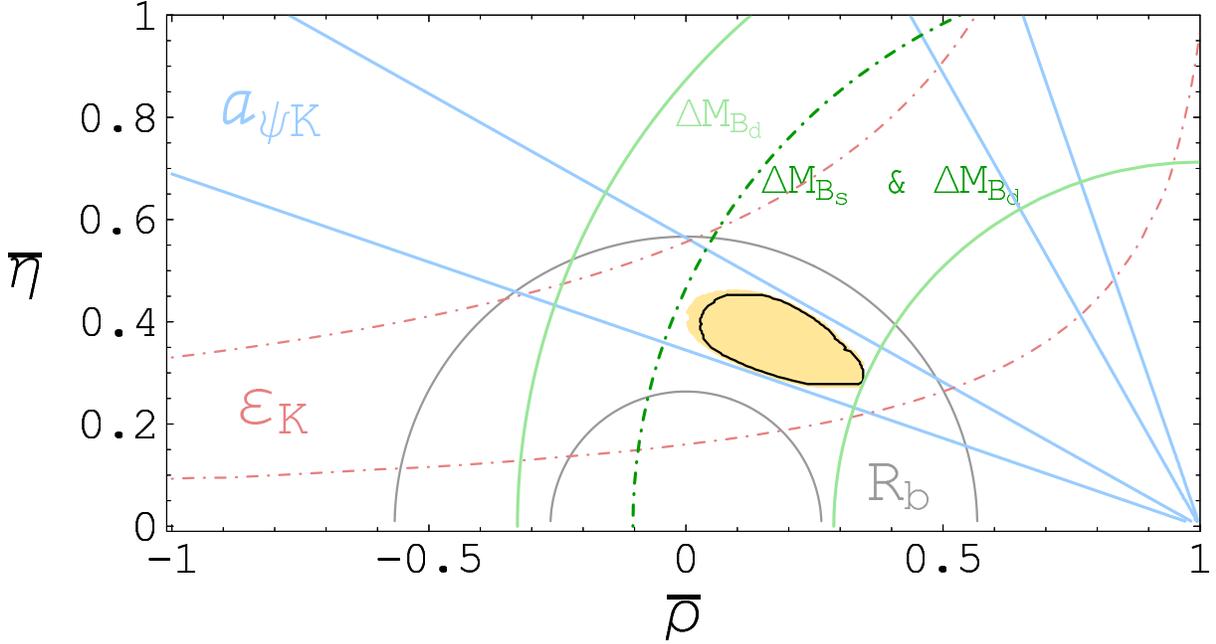}
}
\caption{Constraints in the $\bar\rho - \bar\eta$ plane from the five
         measurements as indicated. Note that the curve labelled 
         as~$\Delta M_{B_s}$ is obtained
         from its 95\%~C.L. lower limit~$14.4$~ps$^{-1}$.
         The fit contour corresponds
         to 95\%~C.L. and the dot shows the best-fit value. The black
         contour shows the impact of the GLSS lower bound 
         $\cos (2\theta) > 0.27$ resulting from the~$C_{\pi\pi}^{+-}$ 
         and $B \to \pi \pi$ branching ratios measurements 
         and using the isospin symmetry.}
\label{fig:SMfit}
\end{figure}

\begin{figure}[htb]
\centerline{
\psfig{width=0.42\textwidth,file=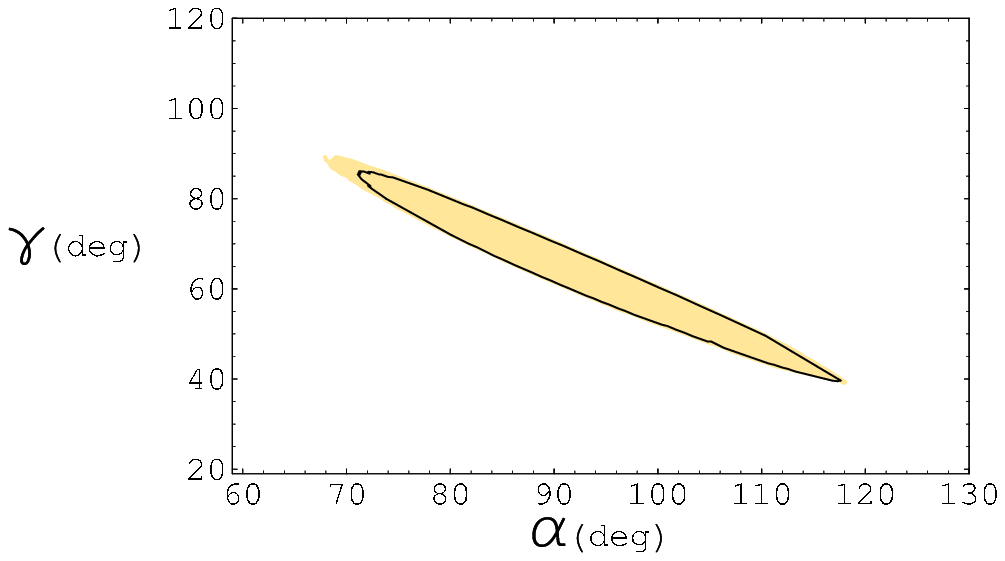}
\quad
\psfig{width=0.56\textwidth,file=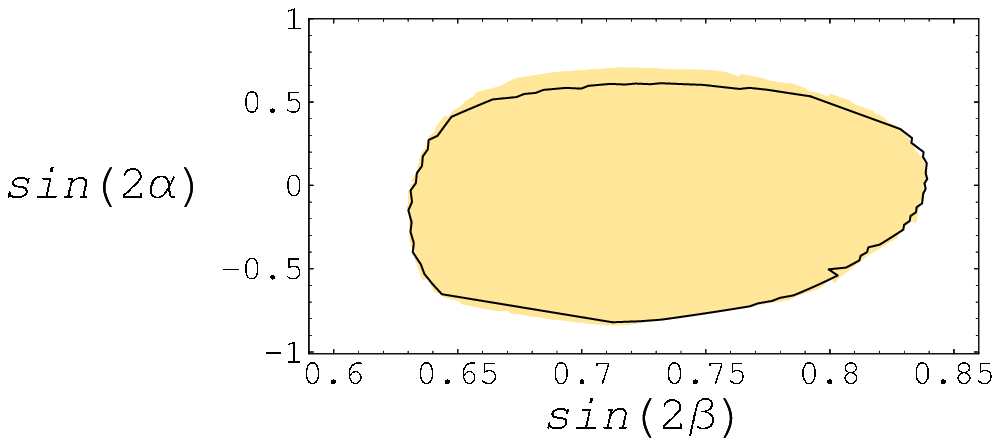}
}
\caption{The 95\%~C.L. correlations between $\alpha - \gamma$ and 
         $\sin (2\alpha) - \sin (2\beta)$ in the~SM. The black contours 
         show the impact of the GLSS lower bound  $\cos (2\theta)> 0.27$ 
         resulting from the~$C_{\pi\pi}^{+-}$ and $B \to \pi \pi$  
         branching ratios measurements and using the isospin symmetry.}
\label{fig:SMfit.alpha-beta-gamma}
\end{figure}

\begin{figure}[htb]
\centerline{
\psfig{width=0.65\textwidth,file=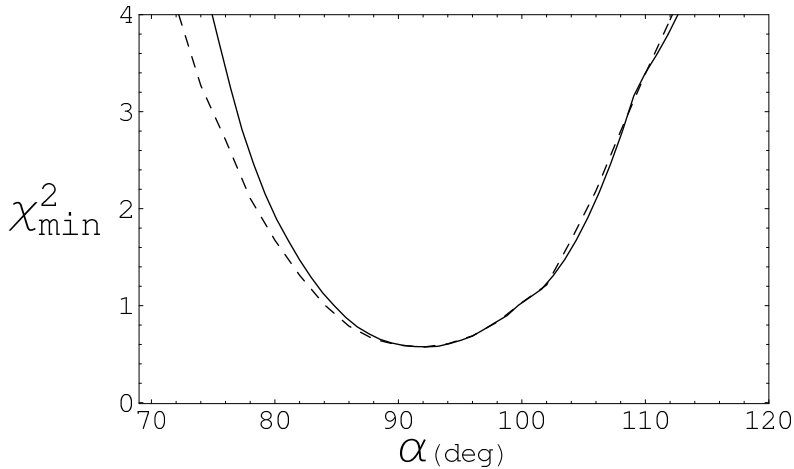}
}
\caption{The $\chi^2_{\rm min}$-distribution as a function of the
         angle~$\alpha$ from the unitarity fits of the CKM 
         parameters in the~SM including the current measurements 
         of~$C_{\pi\pi}^{+-}$ and~$S_{\pi\pi}^{+-}$. The dashed 
         (solid) curve is obtained without (with) taking into 
         account the lower bound $\cos (2\theta) > 0.27$.
}
\label{fig:SMfit.deltalpha}
\end{figure}

We present the output of the fits in Table~\ref{fitvalues}, where 
we show the 68\%~C.L. ranges for the CKM parameters, the angles
of the unitarity triangle, $\Delta M_{B_s}$, $|P_c/T_c|$ 
and~$\delta_c$. Note the enormous ranges for~$|P_c/T_c|$ and~$\delta_c$ 
allowed by the UT fits. The 95\%~C.L. constraints from the five 
individual quantities ($R_b$, $\epsilon_K$, $\Delta M_{B_d}$, 
$\Delta M_{B_s}$, and~$a_{\psi K_S}$) and the resulting fit region 
(the shaded area) are shown in Fig.~\ref{fig:SMfit}.
Further details of this analysis and the discussion of the input 
parameters can be seen elsewhere~\cite{Ali:2003te}. The shaded areas 
in Fig.~\ref{fig:SMfit.alpha-beta-gamma} are the 95\%~C.L. 
correlations between $\alpha - \gamma$ (the left frame) and 
$\sin (2\beta) - \sin (2\alpha)$ (the right frame). 
In Fig.~\ref{fig:SMfit.deltalpha} we show the behaviour of
$\chi^2_{\rm min}$ as a function of the angle~$\alpha$ (the dashed 
curve). The solid curves in all these figures will be explained below.

\begin{figure}[htb]
\centerline{
\psfig{width=0.47\textwidth,file=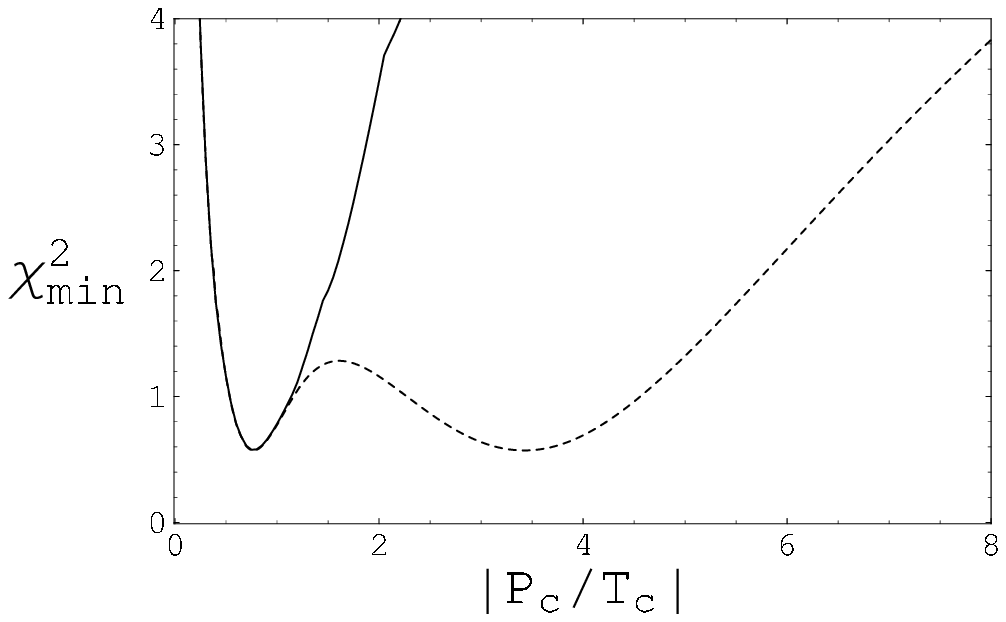}
\quad
\psfig{width=0.47\textwidth,file=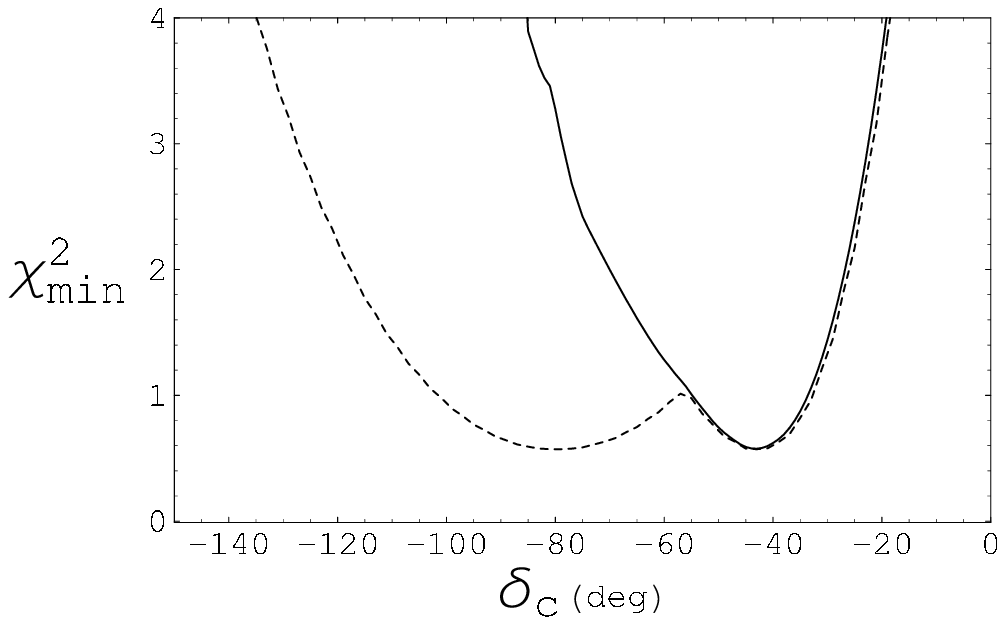}
}
\caption{The $\chi^2_{\rm min}$-distributions as a function 
         of~$|P_c/T_c|$ (left frame) and the strong phase 
         difference~$\delta_c$ (right frame) from the unitarity 
         fits of the CKM parameters in the~SM including the 
         current measurements of~$C_{\pi\pi}^{+-}$ and~$S_{\pi\pi}^{+-}$.  
         The dashed (solid) curve is obtained without (with) 
         taking into account the lower bound $\cos (2\theta) > 0.27$.}
\label{fig:SMfit.deltapt}
\end{figure}

Note that the inclusion of the~$C_{\pi\pi}^{+-}$ and~$S_{\pi\pi}^{+-}$
measurements does not induce any additional constraint on the fits. 
This is because we added two additional terms to the $\chi^2$-function 
with the dependence on two more variables~$|P_c/T_c|$ and~$\delta_c$.
Indeed, for any value of~$\bar \rho$ and~$\bar \eta$, it is always 
possible to choose~$|P_c/T_c|$ and~$\delta_c$ so as to exactly reproduce 
the~$C_{\pi\pi}^{+-}$ and~$S_{\pi\pi}^{+-}$ experimental central values.
Thus, the total~$\chi^2$ of the unitarity triangle fit remains unchanged
as the new measurements do not  
 contribute to the total~$\chi^2$. In the two 
plots presented in Fig.~\ref{fig:SMfit.deltapt} we fix~$|P_c/T_c|$ 
(the left frame) and~$\delta_c$ (the right frame) and minimize 
the $\chi^2$-function with respect to all the other variables. 
In both cases, the absolute minimum of the curve coincides with 
the minimum~$\chi^2$ of the overall fit ($\chi^2_{\rm min} = 0.57$). 
A peculiar feature is the presence of two distinct regions for which 
the overall~$\chi^2$ is very small (the dashed curves). The best-fit 
values for~$|P_c/T_c|$ and~$\delta_c$ are $|P_c/T_c| = 0.77$ and
$\delta_c = -43^\circ$, respectively.
Requiring higher confidence levels for the fits we obtain:
\begin{eqnarray}
|P_c/T_c| \geq 0.23 \quad & \& & \quad 
\delta_c < - 18^\circ \quad 
@ \, 95\%~{\rm C.L.} \, , 
\label{eq:PcTc-dc-95CL} \\
|P_c/T_c| \geq 0.15 \quad & \& & \quad 
\delta_c < - 13^\circ \quad 
@ \, 99\%~{\rm C.L.} \, . 
\label{eq:PcTc-dc-99CL}
\end{eqnarray}

Up to now we have not considered the impact of the isospin-based bounds
in the analysis of the~$B \to \pi \pi $ data on the fits of
the unitarity triangle. In future, the implementation of these bounds
could take the form of the GLSS bound shown in Fig.~\ref{fig:PT-Cos2T}
and the BS bound shown in Fig.~\ref{fig:gam-Spipi}. However, due to the 
proximity of~$S_{\pi\pi}^{+-}$ and $-\sin (2\beta)$ with each other in 
the current data, we will concentrate on implementing the GLSS lower 
bound on $\cos (2\theta)$ resulting from the isospin-based analysis of 
the $B \to \pi \pi$ data presented in the previous section.
A possible implementation of this bound could be undertaken 
by minimizing the $\chi^2$-function and rejecting points for 
which $\cos (2\theta)$ lies below the allowed range. 
Taking $\cos (2\theta) > 0.27$, this analysis results in the black 
contours in Figs.~\ref{fig:SMfit} and~\ref{fig:SMfit.alpha-beta-gamma}. 
Note that a part of the 95\%~C.L. region is now excluded. 
The impact of the $\cos (2\theta)$ 
lower bound on the $\chi^2$-fit for the angle~$\alpha$ is shown 
through the solid curve in  Fig.~\ref{fig:SMfit.deltalpha}. 
This does not change the best-fit value of~$\alpha$ but restricts 
the allowed range of~$\alpha$ by a few degrees at 95\%~C.L. 
The impact of the $\cos (2 \theta)$ lower bound on the fits 
of the unitarity triangle and the correlations $\alpha - \gamma$ 
and $\sin (2\alpha) - \sin (2\beta)$ is currently not great,
but this will change with improved measurements leading to tighter 
constraints on $\cos (2\theta)$, and eventually its measurement. 
The effect of the lower bound on $\cos (2\theta)$ is, however, 
very significant for the allowed value of~$|P_c/T_c|$ and~$\delta_c$. 
This is shown in Fig.~\ref{fig:SMfit.deltapt} through the solid curves 
for~$|P_c/T_c|$ (left frame) and~$\delta_c$ (right frame). While 
the best-fit values of these parameters have not changed, the allowed 
regions are now drastically reduced. Thus, at 68\%~C.L., the allowed 
values are:
\begin{equation}
|P_c/T_c| = 0.77^{+0.58}_{-0.34} , 
\qquad 
\delta_c = (-43^{+14}_{-21})^\circ .  
\label{eq:bestfitparms} 
\end{equation} 
Note that the resulting contours from the analysis of~$S_{\pi\pi}^{+-}$ 
and~$C_{\pi\pi}^{+-}$ do not rely on any model-dependent assumption.
Our results in (\ref{eq:bestfitparms}) can be compared with the analysis 
by Buras {\it et al}.~\cite{Buras:2003dj}, obtained in the SM by 
restricting~$\beta$ and~$\gamma$ in the ranges  
$2\beta = (47 \pm 4)^\circ$ and $\gamma = (65 \pm 7)^\circ$, which yields 
$|P_c/T_c| = 0.49^{+0.33}_{-0.21}$ and $\delta_c = (-43^{+19}_{-23})^\circ$.
The two analyses are compatible with each other though they differ in the
details, in how the exact isospin-relations were imposed in the analysis 
of the data (and also somewhat in the input data). However, we note that 
we have not restricted~$\gamma$ to any range, as it is a fit parameter 
returned by the unitarity fits, but our fit value 
$\gamma = (64 \pm 10)^\circ$ is compatible with the input value 
used by Buras {\it et al}.~\cite{Buras:2003dj}.

\begin{figure}[htb]
\centerline{
\psfig{width=0.65\textwidth,file=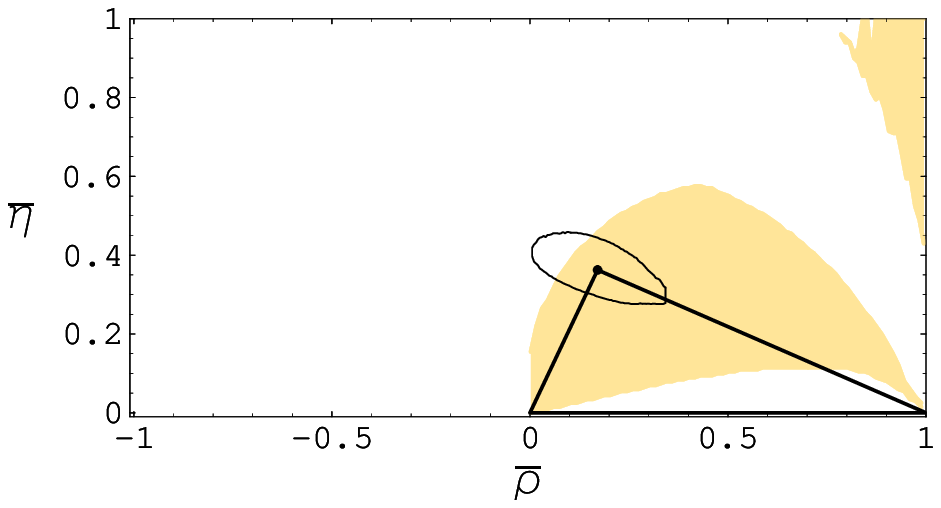}
}
\caption{Constraints in the $\bar\rho - \bar\eta$ plane from the 
         assumed exact measurements of $C_{\pi\pi}^{+-}=-0.46$  
         and~$S_{\pi\pi}^{+-}=-0.74$ and the lower bound 
         $\cos (2\theta) > 0.27$ (the shaded regions). 
         The fit contour corresponds to the 95\%~C.L. unitarity 
         fits without taking into account the~$C_{\pi\pi}^{+-}$ 
         and~$S_{\pi\pi}^{+-}$ measurements.}
\label{fig:SMfit.gedanken}
\end{figure}

Finally, to show the impact of the $\cos (2\theta)$ bound 
on the unitarity triangle in a Gedanken experiment where 
the quantities~$S_{\pi\pi}^{+-}$ and~$C_{\pi\pi}^{+-}$
are assumed to be very precisely measured, we fix~$S_{\pi\pi}^{+-}$ 
and~$C_{\pi\pi}^{+-}$ to their current experimental central 
values~(\ref{eq:Spipi-Cpipi-av-new}) and show the allowed 
region in Fig.~\ref{fig:SMfit.gedanken} resulting from the 
lower bound $\cos (2 \theta) > 0.25$ (the shaded region). 
Note that this results in a constraint in the $\bar \rho - \bar\eta$ 
plane which is very similar to what one gets using a range 
for~$\alpha$. Since the experiments do not measure~$\alpha$, 
but rather~$S_{\pi\pi}^{+-}$ and~$C_{\pi\pi}^{+-}$, this figure 
shows how the eventual~$S_{\pi\pi}^{+-}$ and~$C_{\pi\pi}^{+-}$ 
measurements together with the lower bound on $\cos (2\theta)$ 
gets translated. This represents the strongest possible constraint 
on the profile of the unitarity triangle from the~$S_{\pi\pi}^{+-}$ 
and~$C_{\pi\pi}^{+-}$ measurements that one can get in a 
model-independent way. Of course, this figure itself is only 
illustrative, as the actual constraints will depend on the values
of~$S_{\pi\pi}^{+-}$ and~$C_{\pi\pi}^{+-}$ and the lower bound on 
$\cos (2\theta)$ that will be eventually measured.
\begin{table}[htbp]
\begin{center}
\begin{tabular}{c c c}\hline \hline\\[-2mm]
$\lambda$ & & $\hspace*{5mm}$ $0.2224 \pm 0.002~({\rm fixed})$ 
$\hspace*{5mm}$ \\ 
$\vert V_{cb} \vert $ & & $ (41.2 \pm 2.1) \times 10^{-3} $ \\ 
$\vert V_{ub} \vert$ & & $(3.90 \pm 0.55) \times 10^{-3} $ \\ 
$a_{\psi K_S}$ & & $0.736 \pm 0.049 $ \\ 
$\vert \epsilon_K \vert$ & & $(2.280 \pm 0.13) \times 10^{-3} $ \\ 
$\Delta M_{B_d} $ & & $(0.503 \pm 0.006)~{\rm ps}^{-1}$ \\ 
$\eta_1(m_c(m_c) = 1.30~{\rm GeV})$ & & $1.32 \pm 0.32$ \\ 
$\eta_2 $ & & $0.57 \pm 0.01$ \\ 
$\eta_3$ & & $0.47 \pm 0.05$ \\ 
$m_c (m_c)$ & & $(1.25 \pm 0.10)~{\rm GeV}$ \\
$m_t(m_t)$ & & $(165 \pm 5)~{\rm GeV}$ \\  
$\hat{B}_K$ & & $0.86 \pm 0.15$ \\ 
$f_{B_d} \sqrt{B_{B_d}}$ & & $ (215 \pm 11\pm 15^{+0}_{-23})~{\rm MeV}$ \\ 
$\eta_B$ & & $ 0.55 \pm 0.01$ \\ 
$\xi$ & & $ 1.14 \pm 0.03\pm 0.02^{+0.13}_{-0.0} {}^{+0.03}_{-0.0}$ \\ 
$\Delta M_{B_s} $ & & $ > 14.4~{\rm ps}^{-1}$ at 95\%~C.L. \\[2mm]
\hline\hline
\end{tabular}
\caption{
The input parameters used in the CKM-unitarity fits. 
Their explanation and discussion can be found, for example, 
in Ref.~25. 
}  
\label{inputparms} 
\end{center}
\end{table}
\begin{table}[htbp]
\begin{center}
\begin{tabular}{c c c}\hline \hline\\[-2mm]
$\bar{\rho} $& & $ 0.10 \, \div \, 0.24$\\
$\bar{\eta} $ & & $ 0.32 \, \div \, 0.40 $ \\
$A$ & & $ 0.79  \, \div \,  0.86 $\\[2mm]
$\sin (2\alpha)$ & &  $-0.44 \, \div \, +0.30 $ \\
$\sin (2\beta)$ & &  $ 0.69 \, \div \, 0.78 $\\
$\sin (2\gamma)$ & &  $ 0.50 \, \div \, 0.96$ \\[2mm]
$\alpha$ & &  $ (81 \, \div \, 103)^\circ$\\
$\beta$ & &  $ (21.9 \, \div \, 25.5)^\circ$\\
$\gamma $ & &  $(54 \, \div \, 75)^\circ$\\[2mm]
$\Delta M_{B_s}$ & & $ (16.6 \, \div \, 20.3)~{\rm ps}^{-1}$ \\[2mm] 
$|P_c/T_c|$ & & $ 0.43 \, \div \, 5.3$ \\
$\delta_c$ & & $(-112 \, \div \, -29)^\circ $ \\[2mm]
\hline \hline
\end{tabular}
\caption{The 68\%~C.L. ranges for the CKM-Wolfenstein parameters, 
CP-violating phases, $\Delta M_{B_s}$, $|P_c/T_c|$ and~$\delta_c$ 
from the CKM-unitarity fits.}
\label{fitvalues}
\end{center}
\end{table}
%

%
%
%
%
%
%
%

%
%
%
%

\section{Summary and Concluding Remarks}

We have investigated the impact of the current measurements 
of the time-dependent CP-asymmetry parameters~$S_{\pi\pi}^{+-}$ 
and~$C_{\pi\pi}^{+-}$ in the $B^0_d/\bar{B}^0_d \to \pi^+ \pi^-$ 
decays, reported by the BELLE and BABAR collaborations, on the
CKM parameters. The results of our analysis can be summarized 
as follows.

In the first part of our analysis, we have compared the resulting 
world average~(\ref{eq:Spipi-Cpipi-av-new}) of these measurements 
with the predictions of the specific dynamical 
approaches~\cite{Beneke:1999br,Keum:2000ph} in which the 
quantities~$|P_c/T_c|$ and~$\delta_c$ are estimated. 
We find that, within the~SM, they do not provide a good fit 
of the data. In particular, the estimates of~$|P_c/T_c|$ 
and~$\delta_c$~\cite{Buchalla:2003jr} based on the QCD 
factorization~\cite{Beneke:1999br} are off the mark by more 
than 3~sigma. This was shown in Fig.~\ref{fig:B-pipi} for a 
large enough range of the angle~$\alpha$. The mismatch between 
the data and the QCD-factorization approach~\cite{Beneke:1999br}
can also be studied by calculating the $\chi^2$-function of the
unitarity triangle fit with the estimates by Buchalla and 
Safir~\cite{Buchalla:2003jr} for $|P_c/T_c|$ and~$\delta_c$ 
as an input. The $\chi^2$-minimum of the resulting fit is about~11 
(compared to~$\chi^2_{\rm min}=0.57$, leaving these two parameters
free),
 which corresponds to a probability of about~1\%. Looking more closely  
to localize the source of this discrepancy, our fits show that it 
is the small value of the strong phase difference predicted in the
QCD-factorization approach ($-6^\circ < \delta_c < 24^\circ$ at 
68\%~C.L.) which should be compared to the fit of the data obtained 
by leaving the two parameters free, 
yielding $-64^\circ < \delta_c < -29^\circ$,    
which contributes mainly to the chi square and hence results in the
 poor quality of the fit. 
Since a similar inference also follows for the CP asymmetry 
${\cal A}_{\rm CP} (K^+ \pi^-)$ in the $B^0_d \to K^+ \pi^-$ decay,
where the current measurements yield~\cite{Fry:2003} $(-9.5 \pm 2.9)$\%,
compared to the QCD-factorization prediction~\cite{Beneke:2003zv} 
$(+5\pm 10)$\%, one must conclude, unless data 
change drastically, that this approach grossly underestimates the 
strong interaction phases in the $B \to \pi \pi$ and $B \to K \pi$ 
decays. The competing pQCD approach~\cite{Keum:2000ph} fares 
comparatively somewhat better but predicts a smaller value of~$|P_c/T_c|$
than is required by the current data. The central value of this 
quantity in the estimate by Keum and Sanda~\cite{Keum:2002vi}, 
$|P_c/T_c| = 0.23$, is approximately a factor of two smaller than 
the lowest value of this quantity from the fit range 
$0.43 \le |P_c/T_c| \le 1.35$ at 68\%~C.L., with the best-fit value 
being $|P_c/T_c| = 0.77$. The main message that comes out from this 
part of the analysis is that the QCD-penguin contributions are 
significantly stronger than most of their current phenomenological
estimates, and it remains a theoretical challenge to understand 
this feature of the data.

In the second, and larger part of this paper, we have addressed
the question of how to interpret the measurements of~$S_{\pi\pi}^{+-}$ 
and~$C_{\pi\pi}^{+-}$ in terms of the CKM parameters in a 
model-independent way. We find that leaving the dynamical 
quantities~$|P_c/T_c|$ and~$\delta_c$ as free parameters, the CKM
unitarity fits do not effectively constrain these parameters,
yielding a very large range even at 68\%~C.L.   
As a result of this, the measurements of~$S_{\pi\pi}^{+-}$ 
and~$C_{\pi\pi}^{+-}$ have practically very little impact on the 
unitarity fits of the CKM-Wolfenstein parameters~$\bar \rho$ 
and~$\bar \eta$, and hence on the allowed values of the 
angles~$\alpha$ and~$\gamma$, unless these dynamical quantities 
are bounded. We have reviewed a number of proposals in the 
literature to put isospin-based bounds on the QCD-penguin 
contribution in the $B \to \pi \pi$ decays. Parameterizing it 
in terms of the angle~$2\theta$, introduced by Grossman and Quinn, 
we find that the GLSS lower bound on $\cos (2\theta)$ is the strongest
bound to date, which we have evaluated as $\cos (2\theta) > -0.03$
(propagating the errors on the input quantities) and 
as $\cos (2\theta) > 0.27$ (for the central values).
We have worked out  the consequences of the lower bound  
$\cos (2\theta)> 0.27$ on the profile of the unitarity triangle 
and the angles~$\alpha$ and~$\gamma$. Including the isospin-based
constraint, our best fit values yield: $\alpha = 92^\circ$, 
$\beta = 24^\circ$ and $\gamma = 64^\circ$, with the 68\%~C.L. 
ranges given in Table~\ref{fitvalues}. The corresponding best-fit 
values of the dynamical parameters are found to be $|P_c/T_c| = 0.77$ 
and $\delta_c = -43^\circ$, respectively, with their 68\%~C.L. 
ranges given in Eq.~(\ref{eq:bestfitparms}). With improved data 
expected in the near future, these ranges can be reduced 
significantly, leading to a precise determination of all three 
angles~$\alpha$, $\beta$ and~$\gamma$ of the unitarity triangle.

Of course, at some stage, one has to take into account the 
isospin-breaking corrections.  They originate, in part, from 
the electroweak penguins which are estimated to be numerically 
small~\cite{Deshpande:1994pw,Gronau:1995hn,Fleischer:1995cg},
and this estimate can be put on model-independent 
grounds~\cite{Buras:1998rb,Gronau:1998fn}.  
Moreover, they do not change the bounds obtained above 
for which the closure of the two triangles shown in 
Fig.~\ref{fig:isospin-triangle} was used. However, 
the isospin-breaking corrections may be significant
from the $\pi^0 - \eta - \eta^\prime$ mixing~\cite{Leutwyler:1996tz},
in the presence of which the two isospin triangles used in our analysis 
do not close~\cite{Gardner:1998gz}, leading to quadrilaterals. 
These latter corrections  will have to be accounted for in the 
final determination of the angle~$\alpha$. As the current estimates 
of these corrections are model-dependent~\cite{Gardner:1998gz},
we can not assign at present a quantitative weight to them. 
They will be better determined as and when the individual 
$B^0_d \to \pi^0\pi^0$ and $\bar B^0_d \to \pi^0\pi^0$ 
branching ratios are measured to which we look forward in the future.

\section*{\normalsize Acknowledgements}

We thank James Smith and Andreas H\"ocker for their help with 
the averaging of the current data on~$C_{\pi \pi}^{+-}$ 
and~$S_{\pi \pi}^{+-}$. One of us (A.A.) would also like to
thank Masashi Hazumi for helpful discussions about the BELLE 
data. We are very grateful to David London, Nita and Rahul Sinha 
for pointing out our error in deriving a spurious isospin bound 
in the earlier version of this paper. Helpful communications 
from Andrzej Buras, Robert Fleischer and Dan Pirjol are also 
thankfully acknowledged.
This work is  partially supported by the KEK 
Directorate under a grant from the Japanese Ministry of 
Education, Culture, Sports, Science and Technology.
E.L. and A.Ya.P. acknowledge financial support from the 
Schweizerischer Nationalfonds.

\end{document}